\documentclass[10pt, conference, letterpaper]{IEEEtran}
\usepackage[utf8]{inputenc}
\usepackage[english]{babel}
\usepackage{algorithm}
\usepackage{algorithmicx}
\usepackage{algpseudocode}
\usepackage{amssymb}
\usepackage{array}
\usepackage{arydshln}
\usepackage{blkarray, bigstrut}
\usepackage{bm}
\usepackage{booktabs}
\usepackage{braket}
\usepackage{cite}
\usepackage{cuted}
\usepackage{enumitem}
\usepackage{graphicx}
\usepackage{gensymb}
\usepackage[mathscr]{euscript}
\usepackage[switch]{lineno}
\usepackage{mathtools}
\usepackage{multirow}
\usepackage{multicol}
\usepackage{mathtools}
\usepackage{xfrac}
\usepackage{framed} 
\usepackage[framed]{ntheorem}
\usepackage{pgfplots}
\usepgfplotslibrary{polar}
\usepackage{qtree}
\usepackage{adjustbox}
\usepackage{rotating}
\pgfplotsset{compat=1.13,
    /pgfplots/ybar legend/.style={
    /pgfplots/legend image code/.code={%
       \draw[##1,/tikz/.cd,yshift=-0.25mm]
        (0cm,0cm) rectangle (5pt,2.5mm);},
   },
}
\usepackage{stfloats}
\usepackage{subcaption}
\usepackage{url}
\usepackage{tikz}
\usepackage{tkz-berge}
\usepackage{fixltx2e}
\usepackage[framemethod=TikZ]{mdframed}
\usetikzlibrary{patterns}
\usetikzlibrary{spy}
\usetikzlibrary{shapes, backgrounds,calc, snakes}
\usetikzlibrary{decorations.pathreplacing}
\usetikzlibrary{matrix}
\usepackage{fancybox}

\newframedtheorem{frm-prop}{Property}
\newcommand*\dif{\mathop{}\!\textnormal{\slshape d}}

\definecolor{bblue}{rgb}{0.12392, 0.0490, 0.9588}
\definecolor{sskyblue}{rgb}{0.1529, 0.5882, 0.9216}
\definecolor{ggreen}{rgb}{0.7098, 0.95, 0.40781}
\definecolor{yyellow}{rgb}{0.9765, 0.9804, 0.0784}
\definecolor{dgray}{HTML}{606060}

\definecolor{color0}{HTML}{FF0147}
\definecolor{color1}{HTML}{F400DC}
\definecolor{color2}{HTML}{BA0DFF}
\definecolor{color3}{HTML}{5700E8}
\definecolor{color4}{HTML}{0B03FF}
\definecolor{color5}{HTML}{0957F4}
\definecolor{color6}{HTML}{03B3FF}
\definecolor{color7}{HTML}{08E8DA}
\definecolor{color8}{HTML}{07FF8E}
\definecolor{color9}{HTML}{51FF0A}

\definecolor{p1}{rgb}{1, 0.0667, 0}
\definecolor{p2}{rgb}{1, 0.24, 0}
\definecolor{p3}{rgb}{1, 0.349, 0}
\definecolor{p4}{rgb}{1, 0.490, 0}
\definecolor{p5}{rgb}{1, 0.631, 0}
\definecolor{p6}{rgb}{1, 0.792, 0}
\definecolor{p7}{rgb}{1, 0.933, 0}

\definecolor{ccolor0}{HTML}{FF007D}
\definecolor{ccolor1}{HTML}{760CE8}
\definecolor{ccolor2}{HTML}{0A55FF}
\definecolor{ccolor3}{HTML}{0DB6F4}
\definecolor{ccolor4}{HTML}{00FF76}
\definecolor{ccolor5}{HTML}{6FE80C}
\definecolor{ccolor6}{HTML}{FFDE0A}
\definecolor{ccolor7}{HTML}{FF990A}

\definecolor{mycolor0}{HTML}{e66101}
\definecolor{mycolor1}{HTML}{fdb863}
\definecolor{mycolor2}{HTML}{80cdc1}
\definecolor{mycolor3}{HTML}{018571}
\definecolor{mycolor4}{HTML}{d6604d}
\definecolor{mycolor5}{HTML}{b2182b}

\definecolor{pcolor0}{HTML}{cce7e3}
\definecolor{pcolor1}{HTML}{b3dad4}
\definecolor{pcolor2}{HTML}{99cec6}
\definecolor{pcolor3}{HTML}{80c2b8}
\definecolor{pcolor4}{HTML}{67b6aa}
\definecolor{pcolor5}{HTML}{4daa9c}
\definecolor{pcolor6}{HTML}{349d8d}
\definecolor{pcolor7}{HTML}{1a917f}
\definecolor{pcolor8}{HTML}{018571}
\definecolor{pcolor9}{HTML}{017866}
\definecolor{pcolor10}{HTML}{016a5a}
\definecolor{pcolor11}{HTML}{015d4f}
\definecolor{pcolor12}{HTML}{015044}
\definecolor{pcolor13}{HTML}{014339}
\definecolor{pcolor14}{HTML}{00352d}
\definecolor{pcolor15}{HTML}{002822}
\definecolor{pcolor16}{HTML}{001b17}

\DeclareMathSizes{10pt}{8pt}{6pt}{6pt} 	% Reduce math font size

\usepackage{todonotes}%disable todo notes: [disable]

\newcommand{\fref}[1]{Fig.~\ref{#1}}

\makeatletter
\newcommand\fs@betterruled{%
  \def\@fs@cfont{\bfseries}\let\@fs@capt\floatc@ruled
  \def\@fs@pre{\vspace*{8pt}\hrule height.8pt depth0pt \kern2pt}%
  \def\@fs@post{\kern2pt\hrule\relax}%
  \def\@fs@mid{\kern2pt\hrule\kern2pt}%
  \let\@fs@iftopcapt\iftrue}
\floatstyle{betterruled}
\restylefloat{algorithm}
\makeatother

\def\BibTeX{{\rm B\kern-.05em{\sc i\kern-.025em b}\kern-.08em T\kern-.1667em\lower.7ex\hbox{E}\kern-.125emX}}

\begin{document}
\title{Learning-based Max-Min Fair Hybrid Precoding \\ for mmWave Multicasting}

\author{
\IEEEauthorblockN{Luis F. Abanto-Leon and Gek Hong (Allyson) Sim} 
\IEEEauthorblockA{Secure Mobile Networking (SEEMOO) Lab, Technische Universit\"{a}t Darmstadt, Germany}
\{labanto, asim\}@seemoo.tu-darmstadt.de
}

% Include MAKI ack
\thanks{M. Shell was with the Department
of Electrical and Computer Engineering, Georgia Institute of Technology, Atlanta,
GA, 30332 USA e-mail: (see http://www.michaelshell.org/contact.html).}% <-this % stops a space
\thanks{J. Doe and J. Doe are with Anonymous University.}% <-this % stops a space
\thanks{Manuscript received April 19, 2005; revised August 26, 2015.}

\markboth{Journal of \LaTeX\ Class Files,~Vol.~14, No.~8, August~2015}%
{Shell \MakeLowercase{\textit{et al.}}: Bare Demo of IEEEtran.cls for IEEE Communications Society Journals}

% make the title area
\maketitle

\begin{abstract}
This paper investigates the joint design of hybrid transmit precoder and analog receive combiners for single-group multicasting in millimeter-wave systems. We propose \texttt{LB-GDM}, a low-complexity learning-based approach that leverages gradient descent with momentum and alternating optimization to design \emph{(i)} the digital and analog constituents of a hybrid transmitter and \emph{(ii)} the analog combiners of each receiver. In addition, we also extend our proposed approach to design fully-digital precoders. We show through numerical evaluation that, implementing \texttt{LB-GDM} in either hybrid or digital precoders attains superlative performance compared to competing designs based on semidefinite relaxation. Specifically, in terms of minimum signal-to-noise ratio, we report a remarkable improvement with gains of up to $ 105\% $ and $ 101 \% $ for the fully-digital and hybrid precoders, respectively.  

%This paper investigates the joint design of hybrid transmit precoders and receive combiners for single-group multicasting in millimeter-wave (mmWave) systems. We propose a low complexity LB-GDM, a learning-based approach that leverages gradient descent with momentum and alternating optimization to design \emph{(i)} the digital and analog components of a hybrid transmitter and \emph{(ii)} the combiners of each receiver.  
%Besides, we also extend our proposed approach in the fully-digital precoder design. We show through extensive numerical evaluation that, implementing LB-GDM in either hybrid or digital precoder attain superlative performance compared to competing designs based on semidefinite relaxation (SDR). Specifically, in terms of minimum signal-to-noise ratio (SNR), we report a remarkable improvement with gains of up to $  106.63\% $ and $ 108.63 \% $ higher for the fully-digital and hybrid precoders, respectively.

%Our proposed approach can also be utilized to design fully-digital precoders as an special case. Furthermore, through simulations we corroborate that precoders obtained via LB-GDM attain superlative performance compared to competing designs based on semidefinite relaxation (SDR). Considering the minimum signal--to--noise ratio (SNR) as a performance indicator, we report a remarkable improvement with gains $ X \% $ and $ Y \% $ higher for the fully-digital and hybrid precoders, respectively. 
\end{abstract}

\begin{IEEEkeywords}
max-min fairness, hybrid precoding, multicast, millimeter-wave, learning, semidefinite relaxation.
\end{IEEEkeywords}

\IEEEpeerreviewmaketitle

\section{Introduction}
\label{sec:introduction}
Wireless multicasting has a long-standing record for efficient utilization of spectrum resources to disseminate common information. Looking at the unprecedented growth in number and variety of multicast applications (e.g., high-definition video streaming, mobile video, content distribution in autonomous vehicular networks), multicast is outlined as a key player in emerging 5G millimeter-wave (mmWave) networks to sustain these demands \cite{biason2019:multicast-point-multipoint-directional-mmwave}. With the recent advancements in antenna arrays architectures (e.g., digital-analog designs), particularly for mmWave systems, continuous investigation on beamforming techniques is crucial to ensure high performance. Indeed, a vital aspect to ensure high spectral efficiency lies in the optimal design of the beamformer or precoder. Nevertheless, the optimization problems derived from this context are at best non-convex quadratically constrained quadratic programs (QCQP), which have been proven NP-hard \cite{sidiropoulos2006:transmit-beamforming-physical-layer-multicasting}. Therefore, many ongoing works are devoted to exploring alternative low-complexity schemes that yield near-optimality. 

\subsection{Related work} 
An initial work that addresses the NP-hardness of multicast optimization problems (e.g., quality-of-service (QoS) and max-min fairness (MMF)) in single-group scenarios is \cite{sidiropoulos2006:transmit-beamforming-physical-layer-multicasting}, where non-convex QCQPs are reformulated as semidefinite relaxation (SDR) programs. It is shown that SDR yields an approximate solution that, if feasible, is not necessarily optimum. To find feasible solutions, three types of Gaussian randomization are evaluated. In \cite{tran2014:conic-quadratic-programming-multicasting-arrays}, an iterative algorithm based on second-order conic programming (SOCP) is proposed for the QoS problem in single-group multicasting. The single-group MMF problem is studied in \cite{gopalakrishnan2015:adaptive-algorithms-single-group-multicasting}. Furthermore, the QoS and MMF problems in multi-group multicast contexts are studied in \cite{karipidis2008:qos-maxmin-beamforming-multigroup-multicasting, karipidis2005:transmit-beamforming-multiple-multicast-groups, bornhorst2011:convex-approximation-beamforming-multigroup-multicasting, schad2012:maxmin-transmit-beamforming-multigroup-multicasting, christopoulos2014:multicast-multigroup-beamforming-antenna-power-constraints, demir2015:multigroup-multicast-beamforming-swipt, sadeghi2018:maxmin-precoding-multigroup-multicasting-massive-mimo}.

%Given the affordable hardware cost, low computational complexity, and reduced power consumption of a hybrid precoder, future multi-antenna systems foresee their implementation of hybrid precoders at \textbf{both the infrastructure and mobile devices}. 
%Given the affordable hardware cost, low computational complexity, and small footprint of a hybrid precoder (footprint of what?), future multi-antenna systems foresee the implementation of hybrid precoder at both the infrastructure (e.g., base station and router) and mobile ends. 

The above-mentioned works consider beamforming using fully-digital precoders. In such an architecture, each antenna requires a dedicated baseband and a radio frequency (RF) chain, which is deemed impractical in many multi-antenna systems (e.g., mmWave) due to high design complexity, hardware cost, and power consumption. Consequently, industry and academia scrutinize antenna designs based on a digital-analog (hybrid) architectures which allow the use of a large number of antennas with a limited amount of RF chains \cite{kim2013:mmWave-beamforming-next-generation-communications}. While fully-digital precoders for physical layer multicasting has been widely researched, the design of hybrid precoders remains understudied. The existing literature on hybrid precoding includes investigations on the MMF (in \cite{dai2016:hybrid-precoding-physical-layer-multicasting, sadeghi2018:hybrid-precoding-multigroup-multicasting}) and QoS (in \cite{huang2017:hybrid-beamforming-multicast-transmission-mmwave, demir2016:antenna-selection-hybrid-beamforming-swipt-multigroup-multicasting}) problems for single-group and multi-group multicasting. However, the designs proposed therein are either \emph{(i)} constrained due to simplified premises or \emph{(ii)} unimplementable in the existing multi-antenna hardware, for the following reasons. In \cite{sadeghi2018:hybrid-precoding-multigroup-multicasting}, the propounded solution requires a specially connected network of phase shifters for optimal operation. On the other hand, the proposed scheme in \cite{demir2016:antenna-selection-hybrid-beamforming-swipt-multigroup-multicasting} is restricted to implementations with only four different phase shifts. In \cite{huang2017:hybrid-beamforming-multicast-transmission-mmwave}, the analog phase shifters are replaced by high-resolution lens arrays with adjustable power, thus circumventing the actual problem of phase shift selection. Finally, in \cite{dai2016:hybrid-precoding-physical-layer-multicasting}, it is required to test several codewords in order to design the analog precoder, thus demanding additional memory storage that scales with the number of antennas. 

Our objective is to provide a low-complexity scheme for already available off-the-shelf devices (e.g., TP-Link TALON AD7200), which reckon with a primitive network of phase shifters, limited memory storage, and moderate computational capabilities \cite{palacios2018:adaptive-codebook-beam-training-ieee80211ad}. To address all these requirements, we propose a learning-based scheme that only requires matrix multiplications/additions with controllable complexity and performance that depend on customizable input parameters. Furthermore, in contrast to prior literature on multicasting, we include the design of analog multi-antenna combiners at the receivers.

\subsection{Our contributions}
We design the first learning-based hybrid precoder for single-group multicasting while considering analog multi-antenna receivers. The details of our contributions are summarized as follows: 
\begin{itemize}

\item We investigate the MMF problem subject to power constraints at the transmitter and receivers. Precisely, our solution can handle an arbitrary number of constant-modulus phase shifts for the analog precoder in contrast to the existing designs that only consider a limited number of phase shifts. Moreover, the idea is extended for designing the analog combiners at the receivers.
  
\item Our proposed learning-based scheme has lower complexity than SDR-based approaches. While SDR-based solutions require expensive vector-lifting that expands the variables into higher dimensional spaces, our proposed scheme, namely \texttt{LB-GDM}, only uses matrix multiplications/additions and a number of low-dimensional matrix inversions. Furthermore, the exploration and exploitation phases of our algorithm promote the search for optimal solutions while preventing getting trapped in local optima. Specifically, \texttt{LB-GDM} leverages gradient descent with momentum and alternating optimization.
 
\item We consider analog multi-antenna receivers. We show that, by endowing the receivers with only two antennas, the minimum SNR improves by $ 75.7\% $ compared to omnidirectional receiving patterns (i.e., single antenna case).

\item Since the SDR method in \cite{sidiropoulos2006:transmit-beamforming-physical-layer-multicasting} is only applicable to fully-digital implementations, we propose a novel scheme called \texttt{SDR-C}, capable of handling the constant-modulus constraints of the hybrid precoder and analog receivers. Inspired by \cite{ma2004:semidefinite-relaxation-multiuser-detection-psk}, \texttt{SDR-C} exploits SDR and Cholesky matrix factorization. A similar technique was used by \cite{abanto2019:hybrid-precoding-multigroup-multicasting-mmwave} to solve the QoS problem for multi-group multicasting. We extend the idea in \cite{abanto2019:hybrid-precoding-multigroup-multicasting-mmwave} to the MMF problem.

\item We perform extensive simulations to evaluate the performance of \texttt{LB-GDM} and \texttt{SDR-C} in terms of minimum SNR and spectral efficiency. We provide valuable insights on the fully-digital and hybrid precoders design under various system parameters (i.e., the number of transmit and receive antennas, the number of RF chains, and the number of iterations). We show that \texttt{LB-GDM} substantially outperforms state-of-the-art SDR-based solutions such as \texttt{SDR-C}, achieving up to $ 105.6\% $ and $ 101.4\% $ gains in digital and hybrid precoders, respectively.

\end{itemize}

\section{System Model}
\label{sec:system_model}
We consider a mmWave system where a next generation Node B (gNodeB) serves a set of $ K $ multicast users denoted  by $ \mathcal{K} = \left\lbrace 1, 2, \dots, K \right\rbrace $. The gNodeB is equipped with $ N_\mathrm{tx} $ transmit antennas and $ N^\mathrm{RF}_\mathrm{tx} $ radio frequency (RF) chains, where $ N^\mathrm{RF}_\mathrm{tx} \leq N_\mathrm{tx} $. The downlink signal is represented by $ \mathbf{x} = \mathbf{F} \mathbf{m} s $, where $ \mathbf{F} \in \mathbb{C}^{N_\mathrm{tx} \times N^\mathrm{RF}_\mathrm{tx}} $ and $ \mathbf{m} \in \mathbb{C}^{N^\mathrm{RF}_\mathrm{tx} \times 1} $ are the analog and digital components of the hybrid precoder. The data symbol $ s $ has unit power in average, i.e., $\mathbb{E} \left\lbrace s s^{*} \right\rbrace = 1 $. Every element of the analog precoder is a phase rotation with constant modulus, i.e., $ \left[ \mathbf{F} \right]_{q,r} \in \mathcal{F} = \left\lbrace \sqrt{\delta_\mathrm{tx}}, \dots, \sqrt{\delta_\mathrm{tx}} e^{j \frac{2 \pi \left( L_\mathrm{tx} - 1 \right)}{L_\mathrm{tx}}} \right\rbrace $, where $ q \in \mathcal{Q} = \left\lbrace 1, \dots, N_\mathrm{tx} \right\rbrace $, $ r \in \mathcal{R} = \left\lbrace 1, \dots, N^\mathrm{RF}_\mathrm{tx} \right\rbrace $ and $ L_\mathrm{tx} $ is the number of allowed phase rotation values. Each user is endowed with $ N_\mathrm{rx} \ll N_\mathrm{tx} $ antennas and an analog combiner $ \mathbf{w}_k \in {\mathbb{C}}^{N_\mathrm{rx} \times 1}$ with $ N^\mathrm{RF}_\mathrm{rx} = 1 $, such that $ \left[ \mathbf{w}_k \right] \in \mathcal{W} = \left\lbrace \sqrt{\delta_\mathrm{rx}}, \dots, \sqrt{\delta_\mathrm{rx}} e^{j \frac{2 \pi \left( L_\mathrm{rx} - 1 \right)}{L_\mathrm{rx}}} \right\rbrace $, $ l \in \mathcal{L} = \left\lbrace 1, \dots, N_\mathrm{rx} \right\rbrace $ and $ L_\mathrm{rx} $ is the number of allowed phase rotation possibilities at the receivers. Under the assumption of narrowband flat-fading, the signal received by the $ k $-th user is
% Equation 1
\begin{align} \label{e1}
	y_k = \underbrace{ \mathbf{w}^H_k \mathbf{H}_k \mathbf{F} \mathbf{m} s }_{\text{multicast signal}} 
		  + \underbrace{ \mathbf{w}^H_k \mathbf{n}_k }_{\text{noise}}, 
\end{align}
where $\mathbf{H}_k \in {\mathbb{C}}^{N_\mathrm{rx} \times N_\mathrm{tx}}$ denotes the channel between the $ k $-th user and the gNodeB, whereas $\mathbf{n}_k \sim \mathcal{CN} \left( \mathbf{0}, {\sigma}^2 \mathbf{I} \right) $ denotes additive white Gaussian noise. The SNR at user $ k $ is given by
% Equation 2
\begin{equation} \label{e2}
	\gamma_k = \frac{\left| \mathbf{w}^H_k \mathbf{H}_k \mathbf{F} \mathbf{m} \right|^2} {\sigma^2 \left\| \mathbf{w}_k \right\|^2_2}.
\end{equation}

\section{Problem Formulation}
\label{sec:problem_formulation}
The objective is to design a hybrid precoder that maximizes the minimum SNR among all $ K $ users, subject to power constraints at the transmitter and receiver. We define
% Equation 3
\begin{subequations} \label{e3}
	\begin{align}
	% Objective 3a
	\mathcal{P}^{\text{hyb}}_0: & \max_{
										\substack{ 
											\mathbf{F},
											\mathbf{m},
											\left\lbrace \mathbf{w}_k \right\rbrace^K_{k=1}
										 		}
							  		    } ~ \min_{k \in \mathcal{K}}			 
	& & 
	{
		\frac{ \left| \mathbf{w}^H_k \mathbf{H}_k \mathbf{F} \mathbf{m} \right|^2 }
		     { {\sigma}^2 \left\| \mathbf{w}_k \right\|^2_2 }
	} \label{e3a}
	\\
	\vspace{-0.2cm}
	% Constraint 3b
	& ~~~~~~~~~~~~ \mathrm{s.t.} & & \left\| \mathbf{F} \mathbf{m} \right\|^2_2 = P^{\mathrm{max}}_\mathrm{tx}, \label{e3b}
	\\
	% Constraint 3c
	& & & \left\| \mathbf{F} \right\|^2_\mathrm{F} = 1, \label{e3c}
	\\
	% Constraint 3d
	& & & \left[ \mathbf{F} \right]_{q,r} \in \mathcal{F}, q \in \mathcal{Q}, r \in \mathcal{R}, \label{e3d}
	\\
	% Constraint 3e
	& & & \left\| \mathbf{w}_k \right\|^2_2 = P^{\mathrm{max}}_\mathrm{rx}, k \in \mathcal{K}, \label{e3e}
	\\
	% Constraint 3f
	& & & \left[ \mathbf{w}_k \right]_l \in \mathcal{W}, l \in \mathcal{L}, \forall k \in \mathcal{K}, \label{e3f}
	\end{align}
\end{subequations}
where (\ref{e3b}) restricts the transmit power of the hybrid precoder, (\ref{e3c}) imposes a power normalization on the phase rotations, (\ref{e3d}) enforces every phase rotation  of the analog precoder to be in $ \mathcal{F} $, (\ref{e3e}) restrains the receive power whereas (\ref{e3f}) constrains the phase rotations of the combiners to $ \mathcal{W} $. The constraints (\ref{e3d}) and (\ref{e3f}) denote non-convex feasible sets due to their combinatorial nature. Also, due to parameter coupling, (\ref{e3b}) is non-convex. The objective function (\ref{e3a}) is defined as the ratio of two quadratic expressions, where the numerator exhibits coupling of three parameters. Thus, $ \mathcal{P}^{\text{hyb}}_0 $ is a non-convex problem. Note that (\ref{e3c}) and (\ref{e3e}) can be circumvented as they are only employed to calculate $ \delta_\mathrm{tx} = {1} / {N^\mathrm{RF}_\mathrm{tx} N_\mathrm{tx}} $ and $ \delta_\mathrm{rx} = {P^{\mathrm{max}}_\mathrm{rx}} / {N_\mathrm{rx}} $.

\begin{mdframed}
\emph{Remark:} When $ N_\mathrm{rx} = 1 $, $ \left\lbrace \mathbf{w}_k \right\rbrace^K_{k=1} = 1 $, and $ \mathbf{F}  = \mathbf{I} $, $ \mathcal{P}^{\text{hyb}}_0 $ collapses to the problem investigated in \cite{sidiropoulos2006:transmit-beamforming-physical-layer-multicasting}, which is known to be NP-hard. Since (\ref{e3}) has additional non-convex constraints, $ \mathcal{P}^{\text{hyb}}_0 $ is thus NP-hard as well. Additionally, when $ N_\mathrm{rx} = 1 $ and $ \left\lbrace \mathbf{w}_k \right\rbrace^K_{k=1} = 1 $, $ \mathcal{P}^{\text{hyb}}_0 $ is equivalent to the problem studied in \cite{dai2016:hybrid-precoding-physical-layer-multicasting}. 
\end{mdframed}

\section{Proposed Scheme}
\label{sec:proposed_solution}
In order to solve (\ref{e3}), we adopt an alternating optimization approach that allows us to decouple the unknown parameters $ \mathbf{F} $, $ \mathbf{m} $, and $ \left\lbrace \mathbf{w}_k \right\rbrace^K_{k=1} $. Thus, $ {\mathcal{P}}^{\text{hyb}}_0 $ in (\ref{e3}) is decomposed into three sub-problems $ {\mathcal{P}}^{\text{hyb}}_1 $, $ {\mathcal{P}}^{\text{hyb}}_2 $, and $ {\mathcal{P}}^{\text{hyb}}_3 $ defined in (\ref{e4}), (\ref{e10}), and (\ref{e12}), respectively. Moreover, for each of the sub-problems we propose a learning-based algorithm that leverages gradient descent with momentum, i.e., \texttt{LB-GDM}. Conversely to \cite{werbos1974:beyond-regression-tools-prediction-analysis}, where the momentum term affects the most recent gradient, in our case the momentum is associated with the fittest known solution (at each iteration). Furthermore, we include two parameters, $ N_{\mathrm{xpr}} $ and $ N_{\mathrm{xpt}} $, that control exploration and exploitation of the learning process, respectively. 

\subsection{Optimization of the analog precoder $\mathbf{F}$}
% Algorithm 1
\setlength{\textfloatsep}{5pt}% Remove \textfloatsep
\begin{algorithm} [!t]
	\scriptsize
	%\DontPrintSemicolon
	\textbf{Input:} The precoders $ \mathbf{F}^{(t-1)} $, $ \mathbf{m}^{(t-1)} $ and receive combiners $ \left\lbrace \mathbf{w}^{(t-1)}_k \right\rbrace^K_{k = 1} $ \\
	\textbf{Output:} The analog precoder $ \mathbf{F}^{(t)} $ \\
	\textbf{Execute:} \\ 
	\vspace{0.1cm}
		\begin{tabular}{m{0.2cm} m{7.5cm}}
			1: & Calculate the weights $ c^{(t)}_k, \forall k \in \mathcal{K} $. \\
			2: & Compute $ \nabla J^F = \sum^K_{k = 1} c^{(t)}_k \nabla_{\mathbf{F}} J^{F}_k \slash \left\| \nabla_{\mathbf{F}} J^{F}_k \right\|_{\mathrm{F}} $. \\
			3: & Compute the normalized gradient $ \nabla \tilde{J}^{(t)}_F = \nabla J^F \slash \left\| \nabla J^F \right\|_{\mathrm{F}} $. \\
			4: & Compute $ \mathbf{F}^{(t)} = \mathbf{F}^{(t-1)} + \rho_F \mathbf{F}^{(t-1)}_{\text{best}} + \alpha_F \nabla \tilde{J}^{(t)}_F $. \\
			5: & Project $ \left[ \mathbf{F}^{(t)} \right]_{q,r} \leftarrow \Pi_\mathcal{F} \left[ \mathbf{F}^{(t)} \right]_{q,r} $ onto $ \mathcal{F} $ to satisfy (\ref{e8b}).
		\end{tabular}
	\caption{Optimization of the analog precoder}
	\label{a1}
\end{algorithm}

Assuming that $ \mathbf{m} $ and $ \left\lbrace \mathbf{w}_k \right\rbrace^K_{k=1} $ are known, we optimize $ \mathbf{F} $,
% Equation 4
\begin{subequations} \label{e4}
	\begin{align}
	% Objective 4a
	\mathcal{P}^{\text{hyb}}_1: & \max_{
										\substack{\mathbf{F}}
									} ~ \min_{k \in \mathcal{K}}
	& &									
	{
		\frac{\left| \mathbf{w}^H_k \mathbf{H}_k \mathbf{F} \mathbf{m} \right|^2}{{\sigma}^2 P^{\mathrm{max}}_\mathrm{rx}} 
	} & \label{e4a} 
	\\
	\vspace{-0.2cm}
	% Constraint 4b
	& ~~~~~ \mathrm{s.t.} & & \left\| \mathbf{F} \mathbf{m} \right\|^2_2 = P^{\mathrm{max}}_\mathrm{tx}, & \label{e4b} 
	\\
	% Constraint 4c
	& & & \left[ \mathbf{F} \right]_{q,r} \in \mathcal{F}, q \in \mathcal{Q}, r \in \mathcal{R}. & \label{e4c} 
	\end{align}
\end{subequations}

In order to reduce the number of constraints, we incorporate (\ref{e4b}) into the objective function (\ref{e4a}). Specifically, we replace $ \frac{\left| \mathbf{w}^H_k \mathbf{H}_k \mathbf{F} \mathbf{m} \right|^2}{{\sigma}^2 P^{\mathrm{max}}_\mathrm{rx}} = \psi \frac{\left| \mathbf{w}^H_k \mathbf{H}_k \mathbf{F} \mathbf{m} \right|^2}{\left\| \mathbf{F} \mathbf{m} \right\|^2_2} $, where $ \psi = \frac{P^{\mathrm{max}}_\mathrm{tx}}{{\sigma}^2 P^{\mathrm{max}}_\mathrm{rx}} $. Notice that $ \psi $ can be disregarded as it is constant for all the users. Thus,
% Equation 5
\begin{subequations} \label{e5}
	\begin{align}
	% Objective 5a
	\overline{\mathcal{P}}^{\text{hyb}}_1: & \max_{
													\substack{ \mathbf{F} }
							  		   		      } ~
						 		        \min_{k \in \mathcal{K}}
	& &						 		        
	{
		\frac{ \mathbf{m}^H \mathbf{F}^H \mathbf{H}^H_k \mathbf{w}_k \mathbf{w}^H_k \mathbf{H}_k \mathbf{F} \mathbf{m} }
		{\mathbf{m}^H \mathbf{F}^H \mathbf{F} \mathbf{m}} 
	} & \label{e5a}
	\\
	\vspace{-0.2cm}
	% Constraint 5b
	& ~~~~~ \mathrm{s.t.} & & \left[ \mathbf{F} \right]_{q,r} \in \mathcal{F}, q \in \mathcal{Q}, r \in \mathcal{R}. & \label{e5b}
	\end{align}
\end{subequations}

Instead of approaching (\ref{e5}), we propose to solve the surrogate problem (\ref{e6}), which consists of a weighted sum of all $ \tau^F_k = \frac{ \mathbf{m}^H \mathbf{F}^H \mathbf{H}^H_k \mathbf{w}_k \mathbf{w}^H_k \mathbf{H}_k \mathbf{F} \mathbf{m} } {\mathbf{m}^H \mathbf{F}^H \mathbf{F} \mathbf{m}} $, as shown in (\ref{e6})
% Equation 6
\begin{subequations} \label{e6}
	\begin{align}
	% Objective 6a
	\widehat{\mathcal{P}}^{\text{hyb}}_1: & \max_{ 
													\substack{ \mathbf{F} } 
												 } ~
	& &	
	{
		\sum^K_{k = 1} c_k \frac{ \mathbf{m}^H \mathbf{F}^H \mathbf{H}^H_k \mathbf{w}_k \mathbf{w}^H_k \mathbf{H}_k \mathbf{F} \mathbf{m} }
		{\mathbf{m}^H \mathbf{F}^H \mathbf{F} \mathbf{m}} 
	} & \label{e6a}
	\\
	\vspace{-0.2cm}
	% Constraint 6b
	& ~ \mathrm{s.t.} & & \left[ \mathbf{F} \right]_{q,r} \in \mathcal{F}, q \in \mathcal{Q}, r \in \mathcal{R}, & \label{e6b}
	\end{align}
\end{subequations}
where $ c_k \geq 0 $ denotes the $ k $-th weighting factor. On the other hand, note that $ \tau^F_k $ is upper-bounded by
% Equation 7
\begin{align} \label{e7}
	\begin{split}
		\tau^F_k & \leq \lambda_{\text{max}} \left( \left( \mathbf{F}^H \mathbf{F} \right)^{-1} \mathbf{F}^H \mathbf{H}^H_k \mathbf{w}_k \mathbf{w}^H_k \mathbf{H}_k \mathbf{F} \right) \\
		      & = \underbrace{ \mathbf{w}^H_k \mathbf{H}_k \mathbf{F} \left( \mathbf{F}^H \mathbf{F} \right)^{-1} \mathbf{F}^H \mathbf{H}^H_k \mathbf{w}_k }_{J^{F}_k},
	\end{split}
\end{align}
where $ \lambda_{\text{max}} ( \cdot ) $ extracts the maximum eigenvalue of matrix $ \left( \mathbf{F}^H \mathbf{F} \right)^{-1} \mathbf{F}^H \mathbf{H}^H_k \mathbf{w}_k \mathbf{w}^H_k \mathbf{H}_k \mathbf{F} $. Upon replacing $ \tau^F_k $ in (\ref{e6}) by its upper bound $ J^F_k $, the problem collapses to
% Equation 8
\begin{subequations} \label{e8}
	\begin{align}
	% Objective 8a
	\widetilde{\mathcal{P}}^{\text{hyb}}_1: & \max_{
													\substack{ \mathbf{F} }
												   } ~
	& &											   	
	{
		\sum^K_{k = 1} c_k \mathbf{w}^H_k \mathbf{H}_k \mathbf{F} \left( \mathbf{F}^H \mathbf{F} \right)^{-1} \mathbf{F}^H \mathbf{H}^H_k \mathbf{w}_k,
	} & \label{e8a}
	\\
	\vspace{-0.2cm}
	% Constraint 8b
	& ~ \mathrm{s.t.} & & \left[ \mathbf{F} \right]_{q,r} \in \mathcal{F}, q \in \mathcal{Q}, r \in \mathcal{R}. & \label{e8b}
	\end{align}
\end{subequations}

Since (\ref{e8a}) is an upper bound for (\ref{e6a}), an optimal solution to (\ref{e8}), in general, may not be optimal to (\ref{e6}). Notice that the performance of the system in (\ref{e8}) will be determined by the minimum $ J^{F}_k $, which can be regarded as a utility function of the $ k $-th user. In order to solve (\ref{e8}), we first compute the gradient of $ \sum^K_{k = 1} c_k J^F_k $ to update $ \mathbf{F} $. Then, we scale the modulus of each $ \left[ \mathbf{F} \right]_{q,r} $ and approximate its phase by the closest available option in $ \mathcal{F} $ in order to comply with (\ref{e8b}), as detailed in Algorithm \ref{a1}. The gradient of $ J^F_k $ with respect to $ \mathbf{F} $ is $ \nabla_{\mathbf{F}} J^{F}_k  = \left( \mathbf{I} - \mathbf{F} \mathbf{F}^{\dagger} \right)^T \left( \mathbf{F}^{\dagger} \mathbf{H}^H_k \mathbf{w}_k \mathbf{w}^H_k \mathbf{H}_k \right)^T, $
%% Equation 9
%\begin{align} \label{e9}
%	\begin{split}
%		\nabla_{\mathbf{F}} J^{F}_k  = \left( \mathbf{I} - \mathbf{F} \mathbf{F}^{\dagger} \right)^T \left( \mathbf{F}^{\dagger} \mathbf{H}^H_k \mathbf{w}_k \mathbf{w}^H_k \mathbf{H}_k \right)^T,
%	\end{split}
%\end{align}
where $ \mathbf{F}^{\dagger} = \left( \mathbf{F}^H \mathbf{F} \right)^{-1} \mathbf{F}^H $ (see Appendix for derivation). In \emph{Step~1}, the weights are computed according to $ c^{(t)}_k = \left( 1 + \xi {\left( \gamma^{(t-1)}_{\text{max}} - \gamma^{(t-1)}_k  \right) } \slash { \gamma^{(t-1)}_{\text{max}} } \right)^2 $ for each iteration $ t $, where $ \gamma^{(t)}_k $ is the SNR attained by user $ k $, $ \gamma^{(t)}_{\text{max}} = \max_{k \in \mathcal{K}} \gamma^{(t)}_k $ and $ \xi > 0 $. In \emph{Step~2}, the weighted sum of the unit-power gradients $ { \nabla_{\mathbf{F}} J^{F}_k } \slash { \left\| \nabla_{\mathbf{F}} J^{F}_k \right\|_{\mathrm{F}} } $ is computed. In \emph{Step~3}, the unit-power aggregate gradient $ \nabla \tilde{J}^{(t)}_F $ is obtained. In \emph{Step~4}, the current $ \mathbf{F}^{(t-1)} $ is updated using $ \nabla \tilde{J}^{(t)}_F $. Also, $ \mathbf{F}^{(t)}_{\text{best}} $ represents the best known solution until iteration $ t $, whereas $ \rho_F $ and $ \alpha_F $ are the momentum and learning factors associated to $ \mathbf{F} $, respectively. Finally, \emph{Step~5} enforces (\ref{e8b}). The weights are bounded to $ 1 \leq c^{(t)}_k \leq \left( 1 + \xi \right)^2  $ and increase inversely proportional to the attained SNR $ \gamma^{(t)}_k $. Thus, the gradient of the user with minimum SNR is weighted with the largest $ c^{(t)}_k $, whereas the gradient of the user with maximum SNR is assigned the smallest $ c^{(t)}_k = 1 $. 

\begin{mdframed}
\emph{Remark:} To motivate the connection between (\ref{e5}) and (\ref{e6}), we assume that (\ref{e6}) can be solved iteratively, and in each iteration we are capable of predicting $ k^\star = \arg \min_{k \in \mathcal{K}} \tau^F_k $. Thus, if we assigned binary values $ c_{k^\star} = 1 $ and $ c_{k \neq k^\star} = 0 $ at each iteration instance, we would indirectly be solving a problem closely related to (\ref{e5}), where the minimum SNR is maximized. However, due to the intractability of predicting such $ k^\star $, we propose to simultaneously maximize a subset of the smallest SNRs by considering non-binary positive weights $ c^{(t)}_k $ that can be adapted based on the SNR values (obtained after each iteration), thus controlling the priorities of $ \tau^F_k $ or $ J^F_k $. This proposed approach also facilitates to keep track of several gradients simultaneously, preventing the search from getting trapped in local optima.
\end{mdframed}

\subsection{Optimization of the digital precoder $\mathbf{m}$}
% Algorithm 2
\setlength{\textfloatsep}{5pt}% Remove \textfloatsep
\begin{algorithm} [!t]
	\scriptsize
	%\DontPrintSemicolon
	\textbf{Input:} The precoders $ \mathbf{F}^{(t)} $, $ \mathbf{m}^{(t-1)} $ and receive combiners $ \left\lbrace \mathbf{w}^{(t-1)}_k \right\rbrace^K_{k = 1} $ \\
	\textbf{Output:} The digital precoder $ \mathbf{m}^{(t)} $ \\
	\textbf{Execute:} \\ 
	\vspace{0.1cm}
	\begin{tabular}{m{0.2cm} m{7.5cm}}
		1: & Calculate the weights $ d^{(t)}_k, \forall k \in \mathcal{K} $. \\
		2: & Compute $ \nabla J^M = \sum^K_{k = 1} d^{(t)}_k { \nabla_{\mathbf{m}} J^{M}_k } \slash { \left\| \nabla_{\mathbf{m}} J^{M}_k \right\|_2 } $. \\
		3: & Compute the normalized gradient $ \nabla \tilde{J}^{(t)}_M = { \nabla J^M } \slash { \left\| \nabla J^M \right\|_2 } $. \\
		4: & Compute $ \mathbf{m}^{(t)} = \mathbf{m}^{(t-1)} + \rho_M \mathbf{m}^{(t-1)}_{\text{best}} + \alpha_M \nabla \tilde{J}^{(t)}_M $. \\
		5: & Normalize $ \mathbf{m}^{(t)} \leftarrow \sqrt{ P^{\mathrm{max}}_\mathrm{tx} } {\mathbf{m}^{(t)}} \slash { \left\| \mathbf{F} \mathbf{m}^{(t)} \right\|_2 } $.
	\end{tabular}
	\caption{Optimization of the digital precoder}
	\label{a2}	
\end{algorithm}

When $ \mathbf{F} $ and $ \left\lbrace \mathbf{w}_k \right\rbrace^K_{k=1} $ are known, the problem collapses to
% Equation 10
\begin{subequations} \label{e10}
	\begin{align}
	% Objective 10a
	\mathcal{P}^{\text{hyb}}_2: & \max_{
										\substack{ \mathbf{m} }
									   } ~
								  \min_{k \in \mathcal{K}}
	& &							  
	{
		\left| \mathbf{w}^H_k \mathbf{H}_k \mathbf{F} \mathbf{m} \right|^2
	} & \label{e10a}
	\\
	\vspace{-0.2cm}
	% Constraint 10b
	& ~~~~ \mathrm{s.t.} & & \left\| \mathbf{F} \mathbf{m} \right\|^2_2 = P^{\mathrm{max}}_\mathrm{tx}. & \label{e10b}
	\end{align}
\end{subequations}

Similarly as in (\ref{e5}) and (\ref{e6}), we recast (\ref{e10}) as 
% Equation 11
\begin{subequations} \label{e11}
	\begin{align}
	% Objective 11a
	\widetilde{\mathcal{P}}^{\text{hyb}}_2: & \max_{
													\substack{ \mathbf{m} }
											       } ~
	& &											       	
	{
		\sum^K_{k = 1} d_k \left| \mathbf{w}^H_k \mathbf{H}_k \mathbf{F} \mathbf{m} \right|^2
	} & & \label{e11a}
	\\
	\vspace{-0.2cm}
	% Constraint 11b
	& ~ \mathrm{s.t.} & & \left\| \mathbf{F} \mathbf{m} \right\|^2_2 = P^{\mathrm{max}}_\mathrm{tx}, & & \label{e11b}
	\end{align}
\end{subequations}
where $ d_k $ is the weight corresponding to $ J^M_k = \left| \mathbf{w}^H_k \mathbf{H}_k \mathbf{F} \mathbf{m} \right|^2 $. Compared to $ \widetilde{\mathcal{P}}^{\text{hyb}}_1 $, where an upper bound $ J^M_k $ for $ \tau^F_k $ was derived, finding such a bound by means of the same procedure is not feasible in this case, as it involves computing the inverse of a rank-1 matrix $ M = \mathbf{m}^* \mathbf{m}^T $. Thus, we assume $ J^M_k = \tau^M_k $. $ \widetilde{\mathcal{P}}^{\text{hyb}}_2 $ is iteratively solved employing Algorithm \ref{a2}, where a similar procedure as in Algorithm \ref{a1} is used to compute $ \mathbf{m} $. Moreover, we assume that $ d^{(t)}_k $ are computed in the same fashion as $ c^{(t)}_k $. The gradient of $ J^M_k $ with respect to $ \mathbf{m} $ is $ \nabla_{\mathbf{m}} J^{M}_k = \mathbf{m}^H \mathbf{F}^H \mathbf{H}^H_k \mathbf{w}_k \mathbf{w}_k \mathbf{H}_k \mathbf{F} $. The main difference between Algorithm \ref{a1} and Algorithm \ref{a2} is \emph{Step 5}, which restricts the transmit power to $ P^{\mathrm{max}}_\mathrm{tx} $.
% Algorithm 3
\setlength{\textfloatsep}{1pt}% Remove \textfloatsep
\begin{algorithm} [!t]
	\scriptsize
	%\DontPrintSemicolon
	\textbf{Input:} The precoders $ \mathbf{F}^{(t)} $, $ \mathbf{m}^{(t)} $ and the receive combiner $ \mathbf{w}^{(t-1)}_k $ \\
	\textbf{Output:} The receive combiner $ \mathbf{w}^{(t)}_k $ \\
	\textbf{Execute:} \\ 
	\vspace{0.1cm}
	\begin{tabular}{m{0.2cm} m{7.5cm}}
		1: & Compute $ \nabla_{\mathbf{w}_k} J^W_k $. \\
		2: & Compute $ \nabla_{\mathbf{w}_k} \tilde{J}^{(t)}_W = { \nabla_{\mathbf{w}_k} J^W_k } \slash { \left\| \nabla_{\mathbf{w}_k} J^W_k \right\|_2 } $. \\
		3: & Compute $ \mathbf{w}^{(t)}_k = \mathbf{w}^{(t-1)}_k + \rho_W \mathbf{w}^{(t-1)}_{\text{best},k} + \alpha_W \nabla_{\mathbf{w}_k} \tilde{J}^{(t)}_W $. \\
		4: & Project $ \left[ \mathbf{w}^{(t)}_k \right]_{l} \leftarrow \Pi_\mathcal{W} \left[ \mathbf{w}^{(t)}_k \right]_{l} $ onto $ \mathcal{W} $, $ \forall l \in \mathcal{L} $ to satisfy (\ref{e13b}).
	\end{tabular}
	\caption{Optimization of the $ k $-th combiner}
	\label{a3}
\end{algorithm}

\subsection{Optimization of the combiners $ \mathbf{w}_k $}
Assuming that $ \mathbf{F} $ and $ \mathbf{m} $ are given, we optimize $ \left\lbrace \mathbf{w}_k \right\rbrace^K_{k = 1} $
% Equation 12
\begin{subequations} \label{e12}
	\begin{align}
	% Objective 12a
	\mathcal{P}^{\text{hyb}}_3: & \max_{
										\substack{ 
													\left\lbrace \mathbf{w}_k \right\rbrace^K_{k=1}
											 	 }
									   }~
						   	  	  \min_{k \in \mathcal{K}}
	& &
	{
		\frac{\left| \mathbf{w}^H_k \mathbf{H}_k \mathbf{F} \mathbf{m} \right|^2}
		{{\sigma}^2 \left\| \mathbf{w}_k \right\|^2_2}
	} & \label{e12a}
	\\
	\vspace{-0.2cm}
	% Constraint 12b
	& ~~~~~~~~ \mathrm{s.t.} & & \left[ \mathbf{w}_k \right]_l \in \mathcal{W}, l \in \mathcal{L}, \forall k \in \mathcal{K}. & \label{e12b}
	\end{align}
\end{subequations}

Note that (\ref{e12}) can be decomposed into $ K $ parallel and independent sub-problems, whereby users will adapt their corresponding $ \mathbf{w}_k $ in order to maximize their own SNR. Also, since $ \left\| \mathbf{w}_k \right\|^2_2 $ is an scalar, each sub-problem reduces to
% Equation 13
\begin{subequations} \label{e13}
	\begin{align}
	% Objective 13a
	\widetilde{\mathcal{P}}^{\text{hyb}}_{3,k}: & \max_{
														\substack{ \mathbf{w}_k }
												 	   }
	& &													 	   
	{
		\left| \mathbf{w}^H_k \mathbf{H}_k \mathbf{F} \mathbf{m} \right|^2
	} & \label{e13a}
	\\
	\vspace{-0.2cm}
	% Constraint 13b
	& ~ \mathrm{s.t.} & & \left[ \mathbf{w}_k \right]_l \in \mathcal{W}, l \in \mathcal{L}, & \label{e13b}
	\end{align}
\end{subequations}
$ \forall  k \in \mathcal{K} $. As in $ \widetilde{\mathcal{P}}^{\text{hyb}}_2 $, we assume $ J^W_k = \tau^W_k = \left| \mathbf{w}^H_k \mathbf{H}_k \mathbf{F} \mathbf{m} \right|^2 $. Moreover, each sub-problem in (\ref{e13}) is similar to (\ref{e8}) except that each user optimizes their own utility function $ J^W_k $. We employ Algorithm \ref{a3} to find $ \left\lbrace \mathbf{w}_k \right\rbrace^K_{k=1} $, where $ \nabla_{\mathbf{w}_k} J^W_k = \mathbf{w}^H_k \mathbf{H}_k \mathbf{F} \mathbf{m} \mathbf{m}^H \mathbf{F}^H \mathbf{H}^H_k $.
%\vspace{-5mm}

For completeness, \texttt{LB-GDM} is summarized in Algorithm \ref{a4}. The exploration phase is based on randomization of $ \mathbf{F} $, $ \mathbf{m} $ and $ \left\lbrace \mathbf{w}_k \right\rbrace^K_{k=1} $ ({\footnotesize \texttt{line 17}}). The exploitation phase harnesses $ \mathbf{F}^{(t)}_{\text{best}} $, $ \mathbf{m}^{(t)}_{\text{best}} $, and $ \left\lbrace \mathbf{w}^{(t)}_{\text{best},k} \right\rbrace^K_{k = 1} $ as the momentum terms, which preserve the fittest known solutions until iteration $ t $ and are updated once per exploration instance ({\footnotesize \texttt{line 16}}). On the other hand, $ \mathbf{F}_{\text{opt}} $, $ \mathbf{m}_{\text{opt}} $, and $ \left\lbrace \mathbf{w}_{\text{opt},k} \right\rbrace^K_{k=1} $ retain the fittest solutions after each exploitation instance ({\footnotesize \texttt{line 10}}). These parameters are updated more frequently since they execute a finer scanning of the search space. Further, to refine the potential solutions in this phase, the learning factors $ \alpha_F $, $ \alpha_M $ and $ \alpha_W $are progressively decreased as the exploration phase advances ({\footnotesize \texttt{line 13}}). However, these learning factors are reset to their original values when a new exploration instance begins ({\footnotesize \texttt{line 18}}). A proper balance between exploration and exploitation allows \texttt{LB-GDM} to produce more suitable precoders than SDR.
% Algorithm 4
\begin{algorithm} [!t]
	\scriptsize
	%\DontPrintSemicolon
	\textbf{Initialize:} \\
	\begin{tabular}{m{0.1cm} m{7.9cm}}
	1: & Assign $ \left[ \mathbf{F}^{(0)} \right]_{q,r} \leftarrow \delta $, $ q = \left\lbrace 1, \dots, N_\mathrm{tx} \right\rbrace $, $ r \leftarrow \mod \left( q, N^\mathrm{RF}_\mathrm{tx} \right) + 1 $, \\ 
	   & $ \mathbf{m}^{(0)} \leftarrow \left[ 1 ~ \mathbf{0}_{1 \times (N^\mathrm{RF}_\mathrm{tx} - 1)} \right]^T $, $ \mathbf{w}^{(0)}_k \leftarrow \left[ 1 ~ \mathbf{0}_{1 \times (N_\mathrm{rx} - 1)} \right]^T $, $ \forall k \in \mathcal{K} $. \\
	2: & Assign $ \mathbf{F}_{\text{best}} \leftarrow \mathbf{0} $, $ \mathbf{m}_{\text{best}} \leftarrow \mathbf{0} $ and $ \left\lbrace \mathbf{w}_{\text{best},k} \right\rbrace \leftarrow \mathbf{0} $. \\
	3: & Assign $ \alpha_F \leftarrow \alpha_{F_0} $, $ \alpha_M \leftarrow \alpha_{M_0} $, $ \alpha_W \leftarrow \alpha_{W_0} $. \\
	4: & Assign $ t \leftarrow 0 $, $ \gamma_T \leftarrow 0 $. \\
	\end{tabular} \\
	\textbf{Execute:} \\ 
	\vspace{0.1cm}
	\begin{tabular}{m{0.1cm} m{7.9cm}}
	5: & \textbf{for}  $ i_\mathrm{xpr} = 1, \dots, N_\mathrm{xpr} $ \textbf{do} (exploration phase) \\ 
	6: & ~~ \textbf{for}  $ i_\mathrm{xpt} = 1, \dots, N_\mathrm{xpt} $ \textbf{do} (exploitation phase) \\ 
	7: & ~~~~ Compute $ \mathbf{F}^{(t)} $, $ \mathbf{m}^{(t)} $, $ \left\lbrace \mathbf{w}^{(t)}_k \right\rbrace^K_{k = 1} $ via Algorithms \ref{a1}, \ref{a2}, \ref{a3}. \\
	8: & ~~~~ Find the minimum SNR, $ \gamma_{\text{min}} $, among all users. \\
	9: & ~~~~ \textbf{if} $ \gamma_{\text{min}} \geq \gamma_T $ \\ 
	10: & ~~~~~~ Assign $ \mathbf{F}_{\text{opt}} \leftarrow \mathbf{F}^{(t)} $, $ \mathbf{m}_{\text{opt}} \leftarrow \mathbf{m}^{(t)} $, $ \left\lbrace \mathbf{w}_{\text{opt},k} \right\rbrace^K_{k=1} \leftarrow \left\lbrace \mathbf{w}^{(t)}_k \right\rbrace^K_{k=1} $. \\
	11:  & ~~~~~~ Assign $ \gamma_T \leftarrow \gamma_{\text{min}} $. \\
	12: & ~~~~ \textbf{end if} \\ 
	13: & ~~~~ Update $ \alpha_F \leftarrow 0.98 ~ \alpha_F $, $ \alpha_M \leftarrow 0.98 ~ \alpha_M $, $ \alpha_W \leftarrow 0.98 ~ \alpha_W $. \\
	14: & ~~~~ Increment $ t \leftarrow t + 1 $. \\
	15: & ~~ \textbf{end for} \\
	16: & ~~ Assign $ \mathbf{F}^{(t)}_{\text{best}} \leftarrow \mathbf{F}_{\text{opt}} $, $ \mathbf{m}^{(t)}_{\text{best}} \leftarrow \mathbf{m}_{\text{opt}} $, $ \left\lbrace \mathbf{w}^{(t)}_{\text{best},k} \right\rbrace^K_{k=1} \leftarrow \left\lbrace \mathbf{w}_{\text{opt},k} \right\rbrace^K_{k=1} $. \\
	17: & ~~ Randomize $ \mathbf{F}^{(t)} $, $ \mathbf{m}^{(t)} $ and $ \left\lbrace \mathbf{w}^{(t)}_k \right\rbrace^K_{k=1} $ enforcing (\ref{e3b}) - (\ref{e3f}). \\
	18: & ~~ Assign $ \alpha_F \leftarrow \alpha_{F_0} $, $ \alpha_M \leftarrow \alpha_{M_0} $, $ \alpha_W \leftarrow \alpha_{W_0} $. \\
	19: & \textbf{end for} \\
	\end{tabular} \\
	\caption{Proposed LB-GDM scheme}
	\label{a4}
\end{algorithm}

\section{Simulation Results}
\label{sec:result}
We consider the geometric channel model with $ N_p = 5 $ propagation paths between the transmitter and each user. Also, $ P^{\mathrm{\mathrm{max}}}_\mathrm{tx} = 1 $ ($ 30 $ dBm), $ P^{\mathrm{\mathrm{max}}}_\mathrm{rx} = 0.01 $ ($ 10 $ dBm), $ \sigma^2 = 1 $ ($ 30 $ dBm), while $ \mathcal{F} $ and $ \mathcal{W} $ consist of $ L_\mathrm{tx} = 8 $ and $ L_\mathrm{rx} = 4 $ different phase shifts, respectively. In the following scenarios, we compare the performance of \texttt{LB-GDM} and \texttt{SDR-C} for fully-digital and hybrid precoders in terms of the minimum SNR (among all users) and the spectral efficiency (SE), computed as the sum-capacity of the whole system. We evaluate several configurations of $ N_\mathrm{tx} $, $ N^{\mathrm{RF}}_\mathrm{tx} $, $ N_\mathrm{rx} $, $ N_{\mathrm{xpr}} $, $ N_{\mathrm{xpt}} $, and $ K $. For \texttt{LB-GDM}, we set $ \rho_{F} = \rho_{M} = \rho_{W} = 0.9 $, $ \alpha_{F_0} = 1 $, $ \alpha_{M_0} = 1 $, $ \alpha_{W_0} = 1 $ and vary $ N_\mathrm{xpr} $, $ N_\mathrm{xpt} $ to control the fitness of the solutions. In the case of \texttt{SDR-C}, we control the number of randomizations $ N_{\mathrm{rand}} $. Furthermore, the numerical results show the average over 100 channel realizations.

%\forluis{@Luis: Why do we have different setting for $N_t$? In case A, we have $N_t=15$, this setting is not used in case B, and in case C we use $N_t=20$, which is not used in previous settings. If there is no strong reason for it and we do have results for instance for $N_t = 16$ for all the scenario, I'd fix $N_t$ to be the same in case A and C, which is also use in B. Also the same for $N_r$, it is 3 in Case C.}

\subsection{Impact of exploration ($ N_{\mathrm{xpr}} $) and exploitation ($ N_{\mathrm{xpt}} $)}
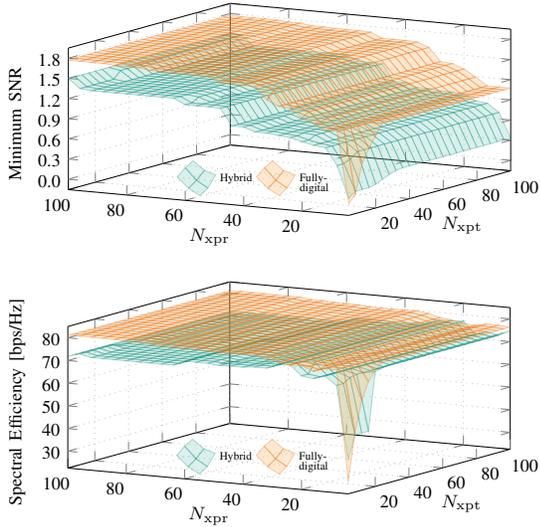
\begin{figure} [!t]
	\vspace{-0.15mm}
	% Figure 1.1: Minimum SNR
	\begin{center}
	\begin{tikzpicture}[scale=0.85]
	\centering
	\begin{axis} 
	[
	height = 4.9cm,
	width = 8.5cm,
	grid = major,
	xmax = 25,
	ymax = 25,
	x label style={at={(0.85,0.045)}, align=center, font=\footnotesize,},
	y label style={at={(0.35,-0.02)}, align=center, font=\footnotesize,},
	z label style={align=center, font=\footnotesize,},
	xlabel = {$ N_{\mathrm{xpt}} $},
	ylabel = {$ N_{\mathrm{xpr}} $},
	zlabel = {Minimum SNR},
	xtick = {5, 10, 15, 20, 25},
	xticklabels = {20, 40, 60, 80, 100},
	x tick label style = {font = \fontsize{8}{9}\selectfont},
	ytick = {5, 10, 15, 20, 25},
	yticklabels = {20, 40, 60, 80, 100},
	y tick label style = {font = \fontsize{8}{9}\selectfont},
	ztick = {0.0, 0.3, 0.6, 0.9, 1.2, 1.5, 1.8},
	zticklabels = {0.0, 0.3, 0.6, 0.9, 1.2, 1.5, 1.8},
	z tick label style = {font = \fontsize{8}{9}\selectfont},
	grid style = {dotted},
	colormap/jet,
	mesh/ordering = x varies,
	mesh/cols = 25,
	mesh/rows = 25,
	view = {-60}{20}, 
	legend columns = 2,
	legend pos = north east,
	legend style={at={(0.23,0.07)},anchor=south west, font=\fontsize{5}{4}\selectfont, text width=0.5cm, text height=0.02cm,text depth=.ex, fill = none, draw = none}
	]
	
	% Hybrid precoder: Nrf = 3 and Ntx = 15
	\addplot3[surf, opacity = 0.3, color = mycolor2, faceted color = mycolor3] table {OSH6MinSNR.txt}; \addlegendentry{Hybrid}
	
	% Digital precoder: Nrf = 15 and Ntx = 15
	\addplot3[surf, opacity = 0.3, color = mycolor1, faceted color = mycolor0] table {OSDMinSNR.txt}; \addlegendentry{Fully-digital}
	\end{axis}
	\end{tikzpicture}
	\end{center}
	
	\vspace{-0.5mm}
	% Figure 1.2: Spectral Efficiency
	\begin{center}
	\begin{tikzpicture}[scale=0.85]
	\begin{axis} 
	[
	height = 4.9cm,
	width = 8.5cm,
	grid=major,
	xmax = 25,
	ymax = 25,
	zmax = 85,
	x label style={at={(0.85,0.045)}, align=center, font=\footnotesize,},
	y label style={at={(0.35,-0.02)}, align=center, font=\footnotesize,},
	z label style={align=center, font=\footnotesize,},
	xlabel = {$ N_{\mathrm{xpt}} $},
	ylabel = {$ N_{\mathrm{xpr}} $},
	zlabel = {Spectral Efficiency [bps/Hz]},
	xtick = {5, 10, 15, 20, 25},
	xticklabels = {20, 40, 60, 80, 100},
	x tick label style = {font = \fontsize{8}{9}\selectfont},
	ytick = {5, 10, 15, 20, 25},
	yticklabels = {20, 40, 60, 80, 100},
	y tick label style = {font = \fontsize{8}{9}\selectfont},
	ztick = {10, 20, 30, 40, 50, 60, 70, 80, 90},
	zticklabels = {10, 20, 30, 40, 50, 60, 70, 80, 90},
	z tick label style = {font = \fontsize{8}{9}\selectfont},
	grid style = {dotted},
	colormap/jet,
	mesh/ordering = x varies,
	mesh/cols = 25,
	mesh/rows = 25,
	view = {-60}{20}, 
	legend columns = 2,
	legend pos = north east,
	legend style={at={(0.23,0.07)},anchor=south west, font=\fontsize{5}{4}\selectfont, text width=0.5cm,text height=0.02cm,text depth=.ex, fill = none, draw = none}
	]

	% Hybrid precoder: Nrf = 3 and Ntx = 15
	\addplot3[surf, opacity = 0.3, color = mycolor2, faceted color = mycolor3] table {OSH6Sum.txt}; \addlegendentry{Hybrid}
	
	% Digital precoder: Nrf = 15 and Ntx = 15
	\addplot3[surf, opacity = 0.3, color = mycolor1, faceted color = mycolor0] table {OSDSum.txt}; \addlegendentry{Fully-digital}
	\end{axis}
	\end{tikzpicture}
	\end{center}
	\vspace{-2mm}
	\caption{Impact of exploration ($N_{\mathrm{xpr}}$) and exploitation ($N_{\mathrm{xpt}}$) phases on the system performance.}
	\label{f1}
	%\vspace{-4mm}
\end{figure}

In this scenario we evaluate the performance of \texttt{LB-GDM} for different values of $ N_{\mathrm{xpr}} $ and $ N_{\mathrm{xpt}} $, under a particular channel realization. We consider $ K = 30 $, $ N_\mathrm{tx} = 15 $, $ N_\mathrm{rx} = 2 $, when $ N_{\mathrm{xpr}} $ and $ N_{\mathrm{xpt}} $ are varied in the range $ \left[ 1, 100 \right] $. For the fully-digital and hybrid precoders, we assume $ N^{\mathrm{RF}}_\mathrm{tx} = N_\mathrm{tx} = 15 $ and $ N^{\mathrm{RF}}_\mathrm{tx} = 6 $, respectively. We observe in Fig. \ref{f1} that the minimum SNR improves for increasing values of $ N_{\mathrm{xpr}} $ and $ N_{\mathrm{xpt}} $ in both precoders. Further, $ N_{\mathrm{xpr}} $ is more relevant than $ N_{\mathrm{xpt}} $ in improving this metric for this particular realization. Nevertheless, both of these phases are important. Exploration is the capability of effectively sampling/scanning the search space to find potentially fitter solutions, whereas exploitation capitalizes on already known solutions to further refine them. By doing so, our proposed \texttt{LB-GDM} avoids getting trapped in local optima. As expected, the fully-digital precoder outperforms its hybrid counterpart due to a larger number of RF chains and less stringent constraints (constant-modulus phase shifts). The former attains a minimum SNR of $ 1.77 $ whereas the latter achieves $ 1.49 $. Besides, the hybrid precoder attains $ 11.5\% $ lower SE than that of the fully-digital precoder. 

\emph{Remark:} While the minimum SNR monotonically increases for both precoders, the SE performance does not exhibit the same behavior. This is because the optimization criterion of \texttt{LB-GDM} is to enhance the minimum SNR (MMF), without considering the spectral efficiency. Nevertheless, the general trend shows that higher $ N_{\mathrm{xpr}} $ and $ N_{\mathrm{xpt}} $ yield SE improvement.
%Figure 1

\vspace{-3mm}
\subsection{Impact of the number of antennas $N_\mathrm{tx}$ and $N_\mathrm{rx}$ }
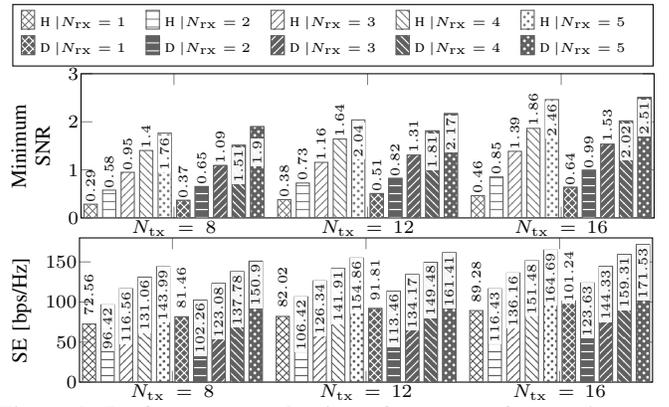
\begin{figure}[!t]
	
	% Figure 2.1: Minimum SNR
	\centering
	\begin{tikzpicture}[inner sep = 0.25mm]
	\begin{axis}
	[
	ybar,
	ymin = 0,
	ymax = 3.0,
	width = 9.2cm,
	height = 35mm,
	bar width = 5.0pt,
	tick align = inside,
	x label style = {align=center, font=\footnotesize,},
	ylabel style = {align=center}, 
	ylabel = Minimum\\SNR,
	y label style={at={(-0.05,0.5)}, font=\footnotesize,},
	nodes near coords,
	every node near coord/.append style={color = black, rotate = 90, anchor = west, font = \fontsize{2}{2}\selectfont},
	nodes near coords align = {vertical},
	symbolic x coords = {$ N_t = 8 $, $ N_t = 12 $, $ N_t = 16 $},
	xticklabels = {$ N_\mathrm{tx} = 8 $, $ N_\mathrm{tx} = 12 $, $ N_\mathrm{tx} = 16 $},
	x tick label style = {text width = 2cm, align = center, font = \fontsize{7}{8}\selectfont},
	y tick label style = {font = \fontsize{7}{8}\selectfont},
	xtick = data,
	enlarge y limits = {value = 0.00, upper},
	enlarge x limits = 0.24,
	legend columns = 5,
	legend pos = north east,
	legend style={at={(-0.12,1.05)},anchor=south west, font=\fontsize{5}{4}\selectfont, text width=1.33cm,text height=0.02cm,text depth=.ex, fill = none}
	]
	
	% Minimum SNR: Hybrid
	\addplot[draw=dgray, pattern color = gray, pattern = crosshatch] coordinates {($ N_t = 8 $, 0.2861) ($ N_t = 12 $, 0.3815) ($ N_t = 16 $, 0.4644)}; \addlegendentry{ H $| N_\mathrm{rx} = 1 $}         
	
	\addplot[draw=dgray, pattern color = gray, pattern = horizontal lines] coordinates {($ N_t = 8 $, 0.5783) ($ N_t = 12 $, 0.7263) ($ N_t = 16 $, 0.8533)}; \addlegendentry{ H $| N_\mathrm{rx} = 2 $}
	
	\addplot[draw=dgray, pattern color = gray, pattern = north east lines] coordinates {($ N_t = 8 $, 0.9494) ($ N_t = 12 $, 1.1561) ($ N_t = 16 $, 1.3856)}; \addlegendentry{ H $| N_\mathrm{rx} = 3 $}
	
	\addplot[draw=dgray, pattern color = gray, pattern = north west lines] coordinates {($ N_t = 8 $, 1.4028) ($ N_t = 12 $, 1.6426) ($ N_t = 16 $, 1.8631)}; \addlegendentry{ H $| N_\mathrm{rx} = 4 $}
	
	\addplot[draw=dgray, pattern color = gray, pattern = crosshatch dots, every node near coord/.append style={fill = white, opacity = 0.95, xshift = -15pt}] coordinates {($ N_t = 8 $, 1.7642) ($ N_t = 12 $, 2.0395) ($ N_t = 16 $, 2.4620)}; \addlegendentry{ H $| N_\mathrm{rx} = 5 $}

	% Minimum SNR: Digital
	\addplot[draw=dgray, fill=dgray, postaction={pattern color = white, pattern = crosshatch}] coordinates {($ N_t = 8 $, 0.3650) ($ N_t = 12 $, 0.5063) ($ N_t = 16 $, 0.6363)}; \addlegendentry{ D $| N_\mathrm{rx} = 1 $}
	
	\addplot[draw=dgray, fill=dgray, postaction={pattern color = white, pattern = horizontal lines}] coordinates {($ N_t = 8 $, 0.6516) ($ N_t = 12 $, 0.8215) ($ N_t = 16 $, 0.9912)}; \addlegendentry{ D $| N_\mathrm{rx} = 2 $}
	
	\addplot[draw=dgray, fill=dgray, postaction={pattern color = white, pattern = north east lines}] coordinates {($ N_t = 8 $, 1.0892) ($ N_t = 12 $, 1.3062) ($ N_t = 16 $, 1.5333)}; \addlegendentry{ D $| N_\mathrm{rx} = 3 $}
	
	\addplot[draw=dgray, fill=dgray, postaction={pattern color = white, pattern = north west lines}, every node near coord/.append style={fill = white, opacity = 0.95, xshift = -15pt}] coordinates {($ N_t = 8 $, 1.5113) ($ N_t = 12 $, 1.8106) ($ N_t = 16 $, 2.0171)}; \addlegendentry{ D $| N_\mathrm{rx} = 4 $}
	
	\addplot[draw=dgray, fill=dgray, postaction={pattern color = white, pattern = crosshatch dots}, every node near coord/.append style={fill = white, opacity = 0.95, xshift = -15pt}] coordinates {($ N_t = 8 $, 1.8980) ($ N_t = 12 $, 2.1690) ($ N_t = 16 $, 2.5073)}; \addlegendentry{ D $| N_\mathrm{rx} = 5 $}
	 
	\end{axis}
	\end{tikzpicture}
	
	% Figure 2.2: Spectral Efficiency
	\centering
	\begin{tikzpicture}[inner sep = 0.25mm]
	\hspace{-0.5mm}
	\begin{axis}
	[
	ybar,
	ymin = 0,
	ymax = 180,
	width = 9.2cm,
	height = 35mm,
	bar width = 5pt,
	tick align = inside,
	x label style={align=center, font=\footnotesize,},
	ylabel style = {text height = 0.4cm, align=center}, 
	ylabel = {SE [bps/Hz]},
	y label style={at={(-0.08,0.5)}, font=\footnotesize,},
	nodes near coords,
	every node near coord/.append style={color = black, rotate = 90, anchor = east, font = \fontsize{2}{2}\selectfont},
	nodes near coords align = {vertical},
	symbolic x coords = {$ N_t = 8 $, $ N_t = 12 $, $ N_t = 16 $},
	xticklabels = {$ N_\mathrm{tx} = 8 $, $ N_\mathrm{tx} = 12 $, $ N_\mathrm{tx} = 16 $},
	x tick label style = {text width = 2cm, align = center, font = \fontsize{7}{8}\selectfont},
	y tick label style = {font = \fontsize{7}{8}\selectfont},
	xtick = data,
	enlarge y limits = {value = 0.00, upper},
	enlarge x limits = 0.24,
	legend columns = 8,
	legend pos = north east,
	legend style={at={(0.14,0.62)},anchor=south west, font=\fontsize{5}{4}\selectfont, text width=0.45cm,text height=0.02cm,text depth=.ex, fill = none}
	]
	
	% Spectral Efficiency: Hybrid
	\addplot[draw=dgray, pattern color = dgray, pattern = crosshatch, every node near coord/.append style={xshift = 20pt}] coordinates {($ N_t = 8 $, 72.5574) ($ N_t = 12 $, 82.0194) ($ N_t = 16 $, 89.2754)}; 
	
	\addplot[draw=dgray, pattern color = dgray, pattern = horizontal lines, every node near coord/.append style={fill = white, opacity = 0.95}] coordinates {($ N_t = 8 $, 96.4155) ($ N_t = 12 $, 106.4234) ($ N_t = 16 $, 116.4297)}; 
	
	\addplot[draw=dgray, pattern color = dgray, pattern = north east lines, every node near coord/.append style={fill = white, opacity = 0.95}] coordinates {($ N_t = 8 $, 116.5648) ($ N_t = 12 $, 126.3406) ($ N_t = 16 $, 136.1595)}; 
	
	\addplot[draw=dgray, pattern color = dgray, pattern = north west lines, every node near coord/.append style={fill = white, opacity = 0.95}] coordinates {($ N_t = 8 $, 131.0592) ($ N_t = 12 $, 141.9118) ($ N_t = 16 $, 151.4839)}; 
	
	\addplot[draw=dgray, pattern color = dgray, pattern = crosshatch dots, every node near coord/.append style={fill = white, opacity = 0.95}] coordinates {($ N_t = 8 $, 143.9925) ($ N_t = 12 $, 154.8558) ($ N_t = 16 $, 164.6866)};

	% Spectral Efficiency: Digital
	\addplot[draw=dgray, fill=dgray, postaction={pattern color = white, pattern = crosshatch}, every node near coord/.append style={fill = white, opacity = 0.95, xshift = 20pt}] coordinates {($ N_t = 8 $, 81.4605) ($ N_t = 12 $, 91.8091) ($ N_t = 16 $, 101.2402)}; 
	
	\addplot[draw=dgray, fill=dgray, postaction={pattern color = white, pattern = horizontal lines}, every node near coord/.append style={fill = white, opacity = 0.95}] coordinates {($ N_t = 8 $, 102.2551) ($ N_t = 12 $, 113.4606) ($ N_t = 16 $, 123.6260)}; 
	
	\addplot[draw=dgray, fill=dgray, postaction={pattern color = white, pattern = north east lines},, every node near coord/.append style={fill = white, opacity = 0.95}] coordinates {($ N_t = 8 $, 123.0847) ($ N_t = 12 $, 134.1744) ($ N_t = 16 $, 144.3314)}; 
	
	\addplot[draw=dgray, fill=dgray, postaction={pattern color = white, pattern = north west lines},, every node near coord/.append style={fill = white, opacity = 0.95}] coordinates {($ N_t = 8 $, 137.7815) ($ N_t = 12 $, 149.4757) ($ N_t = 16 $, 159.3123)}; 
	
	\addplot[draw=dgray, fill=dgray, postaction={pattern color = white, pattern = crosshatch dots},, every node near coord/.append style={fill = white, opacity = 0.95}] coordinates {($ N_t = 8 $, 150.8986) ($ N_t = 12 $, 161.4128) ($ N_t = 16 $, 171.5326)}; 

	\end{axis}
	\end{tikzpicture}
	\vspace{-2mm}
	\caption{Performance evaluation of \texttt{LB-GDM} for varying $ N_\mathrm{tx} $ and $ N_\mathrm{rx} $ in fully-digital (D) and hybrid (H) precoders.}
	\label{f2}
	\vspace{-2mm}
\end{figure}

In this scenario, we evaluate the performance of hybrid and fully-digital precoders based on \texttt{LB-GDM} for a different number of transmit and receive antennas. We consider $ K = 50 $, $ N_\mathrm{tx} = \left\lbrace 8, 12, 16 \right\rbrace $, and $ N_\mathrm{rx} = \left\lbrace 1, 2, 3, 4, 5 \right\rbrace $. For the hybrid precoder, we assume $ N^{\mathrm{RF}}_\mathrm{tx} = 2 $. \fref{f2} depicts the improvement of the minimum SNR when increasing $ N_\mathrm{tx} $ and $ N_\mathrm{rx} $, for both types of precoders. Since the transmit and receive power are limited, endowing users with multiple antennas is beneficial to improve the SNR. In particular, in the fully-digital case, when $ N_\mathrm{tx} = 8 $, the minimum SNR improves from $ 0.37 $ to $ 0.65 $ when the number of receive antenna increases from $ N_\mathrm{rx} = 1 $ to $ N_\mathrm{rx} = 2 $, which essentially indicates a $ 75.7 \% $ gain. Similarly, the gain for the hybrid precoder is $ 100 \% $. We also observe a considerable improvement of the minimum SNR as $ N_\mathrm{tx} $ increases from $ 8 $ to $ 16 $, in which we attain a gain of up to $ 72.9\% $ and $ 58.6\% $ for fully-digital and hybrid precoders, respectively. Further, the SE also achieves $ 25.5\% $ and $ 32.9\% $ gain, for the fully-digital and hybrid precoders, respectively (when $  N_\mathrm{tx} = 8 $, for $ N_\mathrm{rx} = 1 $ and $ N_\mathrm{rx} = 2 $). In general, the hybrid precoder attains a SE at worst $ 11.8\% $ lower than its fully-digital counterpart (for all the cases).  We also observe that with only $ N^{\mathrm{RF}}_\mathrm{tx} = 2 $, the hybrid transmit precoder is at worst $ 25.5\% $ below the optimality attained by the fully-digital in terms of the minimum SNR.

\emph{Remark:} 
This scenario sheds lights on the relevance of reckoning with multiple antennas at the receivers when constrained by power at both ends. Specifically, we obtain improvements up to $ 72.9\% $ and $ 58.6\% $ by increasing the number of receive antennas from $ N_\mathrm{rx} = 1 $ to $ N_\mathrm{rx} = 2 $. On the other hand, in this case where $ N^{\mathrm{RF}}_\mathrm{tx} = 2 $, the complexity of \texttt{LB-GDM} is even more affordable as $ \mathbf{F}^{\dagger} = \left( \mathbf{F}^H \mathbf{F} \right)^{-1} \mathbf{F}^H $ requires no actual inversion of $ \mathbf{F}^H \mathbf{F} $, since a $ 2 \times 2 $ matrix can be inverted directly.
% Figure 2
% Figure 4
\begin{figure}[!t]
	\centering
	% Figure 4.1
	\begin{subfigure}{0.5\textwidth}
	\centering
	\begin{tikzpicture}
	\begin{axis}[
	ybar,
	ymin = 0,
	ymax = 4,
	width = 9.25cm,
	height = 35mm,
	bar width = 5pt,
	tick align = inside,
	x label style={align=center, font=\footnotesize,},
	ylabel = {Min. SNR},
	y label style={at={(-0.075,0.5)}, font=\footnotesize,},
	y tick label style = {font = \fontsize{7}{8}\selectfont},
	%nodes near coords,
	%every node near coord/.append style={color = black, rotate = 90, anchor = west, font = \fontsize{1}{1}\selectfont},
	%nodes near coords align = {vertical},
	symbolic x coords = {$ K = 25 $, $ K = 50 $, $ K = 75 $, $ K = 100 $},
	x tick label style = {text width = 2cm, align = center, font = \fontsize{7}{8}\selectfont},
	xtick = data,
	enlarge y limits = {value = 0.0, upper},
	enlarge x limits = 0.16,
	legend columns = 4,
	legend pos = north east,
	legend style={at={(-0.11,1.08)}, anchor=south west, font=\fontsize{5}{6}\selectfont, text width=17.4mm,text height=0.02cm,text depth=.ex, fill = none},
	% Bottom nodes 
	%% #1: the THRESHOLD after which we switch to a special display.
	nodes near coords bottom/.style={
    	% a new feature since 1.9: allows to place markers absolutely:
    	scatter/position=absolute,
    	close to zero/.style={
        	at={(axis cs:\pgfkeysvalueof{/data point/x},\pgfkeysvalueof{/data point/y})},
    	},
    	big value/.style={
        	at={(axis cs:\pgfkeysvalueof{/data point/x},\pgfkeysvalueof{/data point/y})},
        	color = black, text opacity=1, 
			fill = white, opacity = 0.95,
        	inner ysep=0.5pt, yshift = -13pt, inner sep = 0.17mm
    	},
    	every node near coord/.append style={
      	check for zero/.code={%
        	\pgfmathfloatifflags{\pgfplotspointmeta}{0}{%
            	% If meta=0, make the node a coordinate (which doesn't have text)
            	\pgfkeys{/tikz/coordinate}%
        	}{%
            	\begingroup
            	% this group is merely to switch to FPU locally. Might be
            	% unnecessary, but who knows.
            	\pgfkeys{/pgf/fpu}%
            	\pgfmathparse{\pgfplotspointmeta<#1}%
            	\global\let\result=\pgfmathresult
            	\endgroup
            	% simplifies debugging:
            	%\show\result
            	\pgfmathfloatcreate{1}{1.0}{0}%
            	\let\ONE=\pgfmathresult
            	\ifx\result\ONE
                	% AH : our condition 'y < #1' is met.
                	\pgfkeysalso{/pgfplots/close to zero}%
            	\else
                	% ok, proceed as usual.
                	\pgfkeysalso{/pgfplots/big value}%
            	\fi
        	}
      	},
      	check for zero, 
      	rotate=90, anchor = west, font = \fontsize{0.5}{1}\selectfont,
    	},%
	},%
	nodes near coords,
	% nodes near coords align = {vertical},
	nodes near coords={\pgfmathprintnumber[fixed zerofill,precision=2]{\pgfplotspointmeta}},
	nodes near coords bottom = 1.5,
	]
	
	% Minimum SNR: Hybrid LB-GDM
	\addplot[draw=dgray, fill=dgray, every node near coord/.append style={xshift = -2pt}] coordinates {($ K = 25 $, 3.7781) ($ K = 50 $, 2.2432) ($ K = 75 $, 1.6949) ($ K = 100 $, 1.4265)}; \addlegendentry{ \texttt{LB-GDM} }
	
	% Minimum SNR: Hybrid SDR
	\addplot[draw=dgray, pattern color = dgray, pattern = crosshatch, every node near coord/.append style={xshift = -2pt}] coordinates {($ K = 25 $, 1.8553) ($ K = 50 $, 0.9242) ($ K = 75 $, 0.6479) ($ K = 100 $, 0.4574)}; \addlegendentry{ $ \texttt{SDR-C} \mid 1 $ }
	
	\addplot[draw=dgray, pattern color = dgray, pattern = horizontal lines, every node near coord/.append style={xshift = -2pt}] coordinates {($ K = 25 $, 1.7617) ($ K = 50 $, 1.0072) ($ K = 75 $, 0.7467) ($ K = 100 $, 0.5616)}; \addlegendentry{ $ \texttt{SDR-C} \mid 10 $ }
	
	\addplot[draw=dgray, pattern color = dgray, pattern = north east lines, every node near coord/.append style={xshift = -2pt}] coordinates {($ K = 25 $, 1.9807) ($ K = 50 $, 1.1198) ($ K = 75 $, 0.7963) ($ K = 100 $, 0.6188)}; \addlegendentry{ $ \texttt{SDR-C} \mid 50 $ }
	
	\addplot[draw=dgray, pattern color = dgray, pattern = north west lines, every node near coord/.append style={xshift = -2pt}] coordinates {($ K = 25 $, 1.9335) ($ K = 50 $, 1.1349) ($ K = 75 $, 0.8013) ($ K = 100 $, 0.6176)}; \addlegendentry{ $ \texttt{SDR-C} \mid 100 $ }
	
	\addplot[draw=dgray, pattern color = dgray, pattern = crosshatch dots, every node near coord/.append style={xshift = -2pt}] coordinates {($ K = 25 $, 2.1099) ($ K = 50 $, 1.2826) ($ K = 75 $, 0.8650) ($ K = 100 $, 0.7010)}; \addlegendentry{ $ \texttt{SDR-C} \mid 500 $ }
	
	\addplot[draw=dgray, pattern color = dgray, pattern = grid, every node near coord/.append style={xshift = -2pt}] coordinates {($ K = 25 $, 2.0632) ($ K = 50 $, 1.3418) ($ K = 75 $, 0.8081) ($ K = 100 $, 0.7073)}; \addlegendentry{ $ \texttt{SDR-C} \mid 1000 $ }	             
	            
	\end{axis}
	\end{tikzpicture}
	\vspace{-2mm}
	\caption{Hybrid precoder}
	\label{f3a}
	\end{subfigure}
	
	\vspace{1mm}
	% Figure 4.2
	\begin{subfigure}{0.5\textwidth}
	\centering
	\begin{tikzpicture}
	\begin{axis}[
	ybar,
	ymin = 0,
	ymax = 4.5,
	width = 9.25cm,
	height = 35mm,
	bar width = 5pt,
	tick align = inside,
	x label style={align=center, font=\footnotesize,},
	ylabel = {Min. SNR},
	y label style={at={(-0.075,0.5)}, font=\footnotesize,},
	y tick label style = {font = \fontsize{7}{8}\selectfont},
	%nodes near coords,
	%every node near coord/.append style={color = black, rotate = 90, anchor = west, font = \fontsize{1}{1}\selectfont},
	%nodes near coords align = {vertical},
	symbolic x coords = {$ K = 25 $, $ K = 50 $, $ K = 75 $, $ K = 100 $},
	x tick label style = {text width = 2cm, align = center, font = \fontsize{7}{8}\selectfont},
	xtick = data,
	enlarge y limits = {value = 0.0, upper},
	enlarge x limits = 0.16,
	legend columns = 4,
	legend pos = north east,
	legend style={at={(-0.11,1.08)}, anchor=south west, font=\fontsize{5}{6}\selectfont, text width=17.4mm,text height=0.02cm,text depth=.ex, fill = none},
	% Bottom nodes 
	%% #1: the THRESHOLD after which we switch to a special display.
	nodes near coords bottom/.style={
    	% a new feature since 1.9: allows to place markers absolutely:
    	scatter/position=absolute,
    	close to zero/.style={
        	at={(axis cs:\pgfkeysvalueof{/data point/x},\pgfkeysvalueof{/data point/y})},
    	},
		big value/.style={
        	at={(axis cs:\pgfkeysvalueof{/data point/x},\pgfkeysvalueof{/data point/y})},
        	color = black, text opacity=1, 
			fill = white, opacity = 0.95,
        	inner ysep=0.5pt, yshift = -15pt, inner sep = 0.17mm
    	},
    	every node near coord/.append style={
      	check for zero/.code={%
        	\pgfmathfloatifflags{\pgfplotspointmeta}{0}{%
            	% If meta=0, make the node a coordinate (which doesn't have text)
            	\pgfkeys{/tikz/coordinate}%
        	}{%
            	\begingroup
            	% this group is merely to switch to FPU locally. Might be
            	% unnecessary, but who knows.
            	\pgfkeys{/pgf/fpu}%
            	\pgfmathparse{\pgfplotspointmeta<#1}%
            	\global\let\result=\pgfmathresult
            	\endgroup
            	% simplifies debugging:
            	%\show\result
            	\pgfmathfloatcreate{1}{1.0}{0}%
            	\let\ONE=\pgfmathresult
            	\ifx\result\ONE
                	% AH : our condition 'y < #1' is met.
                	\pgfkeysalso{/pgfplots/close to zero}%
            	\else
                	% ok, proceed as usual.
                	\pgfkeysalso{/pgfplots/big value}%
            	\fi
        	}
      	},
      	check for zero, 
      	font=\footnotesize, 
      	rotate=90, anchor = west, font = \fontsize{0.5}{1}\selectfont,
    	},%
	},%
	nodes near coords,
	% nodes near coords align = {vertical},
	nodes near coords={\pgfmathprintnumber[fixed zerofill,precision=2]{\pgfplotspointmeta}},
	nodes near coords bottom = 1.5,
	]
	
	% Minimum SNR: Digital LB-GDM
	\addplot[fill = dgray,  inner sep=1pt] coordinates {($ K = 25 $, 4.2388) ($ K = 50 $, 2.5985) ($ K = 75 $, 1.8020) ($ K = 100 $, 1.4601)};  \addlegendentry{ \texttt{LB-GDM} }
	
	% Minimum SNR: Digital SDR
	\addplot[draw=dgray, pattern color = dgray, pattern = crosshatch, inner sep=1pt, anchor = west] coordinates {($ K = 25 $, 1.8508) ($ K = 50 $, 0.6161) ($ K = 75 $, 0.3691) ($ K = 100 $, 0.2364)}; \addlegendentry{ $ \texttt{SDR-C} \mid 1 $ }
	
	\addplot[draw=dgray, pattern color = dgray, pattern = horizontal lines, inner sep=1pt] coordinates {($ K = 25 $, 3.0067) ($ K = 50 $, 1.2528) ($ K = 75 $, 0.7075) ($ K = 100 $, 0.4537)}; \addlegendentry{ $ \texttt{SDR-C} \mid 10 $ }
	
	\addplot[draw=dgray, pattern color = dgray, pattern = north east lines, inner sep=1pt] coordinates {($ K = 25 $, 3.1200) ($ K = 50 $, 1.3604) ($ K = 75 $, 0.8172) ($ K = 100 $, 0.5572)}; \addlegendentry{ $ \texttt{SDR-C} \mid 50 $ }
	
	\addplot[draw=dgray, pattern color = dgray, pattern = north west lines, inner sep=1pt] coordinates {($ K = 25 $, 3.1777) ($ K = 50 $, 1.4453) ($ K = 75 $, 0.8097) ($ K = 100 $, 0.6785)}; \addlegendentry{ $ \texttt{SDR-C} \mid 100 $ }
	
	\addplot[draw=dgray, pattern color = dgray, pattern = crosshatch dots, inner sep=1pt] coordinates {($ K = 25 $, 3.1819) ($ K = 50 $, 1.3886) ($ K = 75 $, 0.8474) ($ K = 100 $, 0.6646)}; \addlegendentry{ $ \texttt{SDR-C} \mid 500 $ } 
	
	\addplot[draw=dgray, pattern color = dgray, pattern = grid, inner sep=1pt] coordinates {($ K = 25 $, 3.2951) ($ K = 50 $, 1.4302) ($ K = 75 $, 0.8812) ($ K = 100 $, 0.7087)}; \addlegendentry{ $ \texttt{SDR-C} \mid 1000 $ }
	
	\end{axis}
	\end{tikzpicture}
	\vspace{-2mm}
	\caption{Fully-digital precoder}
	\label{f3b}
	\end{subfigure}
	\vspace{-2mm}
	\caption{Performance comparison between \texttt{LB-GDM} and \texttt{SDR-C} in terms of the minimum SNR.}
	\label{f3}
\end{figure}
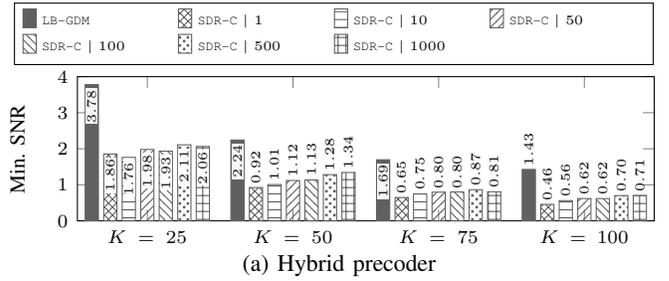
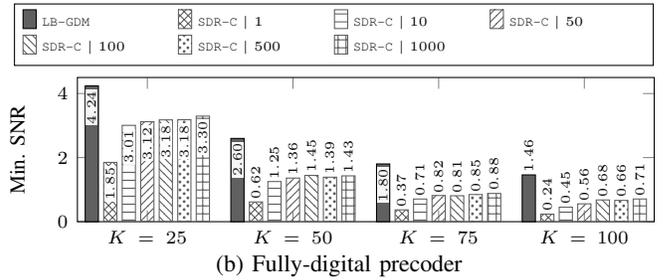

\subsection{Performance comparison with an SDR-based scheme}
We compare the performance of \texttt{LB-GDM} and \texttt{SDR-C}, when implemented in fully-digital and hybrid precoders. We consider $ N_\mathrm{tx} = 20 $, $ N_\mathrm{rx} = 3 $, with a wide range of users $ K = \left\lbrace 25, 50, 75, 100 \right\rbrace $. For the hybrid precoder $ N^{\mathrm{RF}}_\mathrm{tx} = 6 $, whereas for the fully-digital counterpart $ N^\mathrm{RF}_\mathrm{tx} = N_\mathrm{tx} $. For \texttt{LB-GDM}, we assume that $ N_{\mathrm{xpt}} = N_{\mathrm{xpr}} = 120 $. For \texttt{SDR-C}, the number of randomizations are $ N_{\mathrm{rand}} = \left\lbrace 1, 10, 50, 100, 500, 1000 \right\rbrace $. To ensure a fair comparison, we refine the solutions of \texttt{SDR-C} by optimizing sequentially $ \mathbf{F} $, $ \mathbf{m} $, and $ \left\lbrace \mathbf{w}_k \right\rbrace^K_{k = 1} $ over $ N^{{\mathrm{SDR}}}_{\mathrm{iter}} = 3 $ iterations. In each iteration, $ N_{\mathrm{rand}} $ randomizations are evaluated. Fig. \ref{f3} depicts a notable improvement of \texttt{LB-GDM} over \texttt{SDR-C} in both fully-digital (see Fig. \ref{f3b}) and hybrid (see Fig. \ref{f3a}) implementations, for all $ K $. Specifically, the \texttt{SDR-C} results are shown in the format $ \left\langle \texttt{SDR-C} \mid N_\mathrm{rand} \right\rangle $. We observe a more prominent improvement for larger $ K $. For instance, in the case of the fully-digital precoder, when $ K = 50 $, the minimum SNR obtained by \texttt{LB-GDM} is $ 79.3\% $ higher than that of \texttt{SDR-C} although a wide range of $ N_{\mathrm{rand}} $ were tested. The gain is even higher (i.e. $ 105.6\% $) for $ K = 100 $. We observe a similar trend for \texttt{LB-GDM}-based hybrid precoder, with gains of up to $ 101.4\% $. 

\section{Discussion}
\noindent{{\bf\texttt{\small{SDR-C}}:} This scheme is based on the approach in \cite{abanto2019:hybrid-precoding-multigroup-multicasting-mmwave}, where the QoS problem is researched. We extended the approach therein for the MMF problem. In this paper, \texttt{SDR-C} solves the sub-problems $ {\mathcal{P}}^{\text{hyb}}_1 $, $ {\mathcal{P}}^{\text{hyb}}_2 $, $ {\mathcal{P}}^{\text{hyb}}_3 $ in alternate manner over $ N^{{\mathrm{SDR}}}_{\mathrm{iter}} = 3 $ iterations. The initialization of $ \mathbf{m} $ and $ \left\lbrace \mathbf{w}_k \right\rbrace^K_{k=1} $ are the same as for \texttt{LB-GDM} (see \texttt{line 1} of Algorithm \ref{a4}). The SDR-C scheme is discussed in Appendix \ref{xB}.

\noindent{{\bf{\small{Optimality}}:} The proposed schemes, \texttt{LB-GDM} and \texttt{SDR-C}, cannot ensure global optimality. However, by observing Fig. \ref{f1} and Fig. \ref{f3} we corroborate that the approaches converge to a local optima for increasing $ N_\mathrm{xpr} $, $ N_\mathrm{xpt} $ or $ N_\mathrm{rand} $. 

\noindent{{\bf{\small{Impact of number of constraints}}:} It is well known that the optimality-gap of SDR degrades with increasing number of constraints (i.e., number of users $ K $). As a result, we observe that for large $ K $, the performance difference between \texttt{LB-GDM} and \texttt{SDR-C} increases, which indicates that \texttt{LB-GDM} is more robust and less sensitive to the number of constraints.

\section{Conclusion}
\label{sec:conclusion}
In this paper, we investigated the design of fully-digital and hybrid precoders for single-group multicasting using a learning-based scheme. With the aim of maximizing the minimum SNR, our proposed low-complexity \texttt{LB-GDM} uses only matrix multiplications/additions and low-dimensional matrix inversion operations. We compare the performance of precoders based on \texttt{SDR-C} and \texttt{LB-GDM} under diverse simulation settings. The numerical results show a substantial gain, where \texttt{LB-GDM} outperforms \texttt{SDR-C} by up to $ 105.6\% $ and $ 101.4\% $ for digital and hybrid precoders, respectively. In addition, we demonstrate the importance of incorporating more receive antennas, where we achieve $ 75.7 \% $ and $ 100 \% $ gains in terms of the minimum SNR by increasing the number of receive antennas from one to two. 

%In this paper, we investigated the design of fully-digital and hybrid precoders for single-group multicasting using a learning-based scheme. With the aim to maximize the minimum SNR, our low-complexity learning-based scheme uses only matrix multiplications/additions and several low-dimensional matrix inversion operations. Furthermore, we leverage the exploration and exploitation phases to obtain close-to-optimal performance and prevent getting trapped at the local optima (as experienced by SDR-based approach). We compare the performance of the state-of-the-art SDR-based schemes and our proposed LB-GDM-based precoders under diverse simulation settings. The numerical results show substantial gain where LB-GDM outperforms SDR by up to $ 108.6\% $ and $ 101.4\% $  for digital and hybrid precoders, respectively. In addition, we demonstrate the importance of incorporating more receive antennas where we achieve $ 144.4 \% $ gain in terms of minimum SNR by just increasing the number of receive antennas to two. 

\section{Acknowledgment}
This research was in part funded by the Deutsche Forschungsgemeinschaft (DFG) within the B5G-Cell project as part of the SFB 1053 MAKI.
\begin{appendices}
\setcounter{equation}{0}
\renewcommand{\theequation}{A.\arabic{equation}}
\section{Gradient of $ J^F_k $ in Algorithm \ref{a1}} \label{xA}

Let us define $ \mathbf{u} = \mathbf{F}^H \mathbf{H}^H_k \mathbf{w}_k $ and $ \mathbf{Y} = \mathbf{F}^H \mathbf{F} $. Then, the following differentials are computed: $ \dif \mathbf{u} = \dif \mathbf{F}^H \mathbf{H}^H_k \mathbf{w}_k $, $ \dif \mathbf{Y} = \mathbf{F}^H \dif \mathbf{F} $, $ \dif \mathbf{u}^H = \mathbf{w}^H_k \mathbf{H}_k \dif \mathbf{F} $ and $ \dif \mathbf{Y}^{-1} = - \mathbf{Y}^{-1} \dif \mathbf{Y} \mathbf{Y}^{-1} $. Thus, the differential of $ J^{F}_k = \mathbf{u}^H \mathbf{Y}^{-1} \mathbf{u} $ is given by
\begin{align*}
		\dif J^{F}_k = & \left( \dif \mathbf{u}^H \right) \mathbf{Y}^{-1} \mathbf{u} + \mathbf{u}^H \left( \dif \mathbf{Y}^{-1} \right) \mathbf{u} + \mathbf{u}^H \mathbf{Y}^{-1} \left( \dif \mathbf{u} \right) \\
					= & \left( \mathbf{w}^H_k \mathbf{H}_k \dif \mathbf{F} \right) \mathbf{Y}^{-1} \mathbf{u}  - \mathbf{u}^H \left( \mathbf{Y}^{-1} \dif \mathbf{Y} \mathbf{Y}^{-1} \right) \mathbf{u} \\
					= & \left( \mathbf{w}^H_k \mathbf{H}_k \dif \mathbf{F} \right) \mathbf{Y}^{-1} \mathbf{u}  - \mathbf{u}^H \left( \mathbf{Y}^{-1} \mathbf{F}^H \dif \mathbf{F} \mathbf{Y}^{-1} \right) \mathbf{u} \\
					= & ~ \mathrm{Tr} \left\lbrace \mathbf{Y}^{-1} \mathbf{u} \mathbf{w}^H_k \mathbf{H}_k \dif \mathbf{F} \right\rbrace - \mathrm{Tr}\left\lbrace \mathbf{Y}^{-1} \mathbf{u} \mathbf{u}^H \mathbf{Y}^{-1} \mathbf{F}^H \dif \mathbf{F} \right\rbrace \\
					= & ~ \mathrm{Tr} \left\lbrace \left( \mathbf{Y}^{-1} \mathbf{u} \mathbf{w}^H_k \mathbf{H}_k - \mathbf{Y}^{-1} \mathbf{u} \mathbf{u}^H \mathbf{Y}^{-1} \mathbf{F}^H \right) \dif \mathbf{F} \right\rbrace 
\end{align*}

The Frobenius inner product of two matrices $ \mathbf{P} $ and $ \mathbf{Q} $ is defined as $ \mathbf{P} : \mathbf{Q} \equiv \mathrm{Tr} \left\lbrace \mathbf{P}^T \mathbf{Q} \right\rbrace $. Thus, $ \dif J^{F}_k = \left( \mathbf{Y}^{-1} \mathbf{u} \mathbf{w}^H_k \mathbf{H}_k - \mathbf{Y}^{-1} \mathbf{u} \mathbf{u}^H \mathbf{Y}^{-1} \mathbf{F}^H \right)^T : \dif \mathbf{F}. $ Upon replacing $ \mathbf{u} $ in the expression above, we obtain
\begin{align} 
	\begin{split}
		\nabla_{\mathbf{F}} J^{F}_k = \left( \mathbf{I} - \mathbf{F} \mathbf{F}^{\dagger} \right)^T \left( \mathbf{F}^{\dagger} \mathbf{H}^H_k \mathbf{w}_k \mathbf{w}^H_k \mathbf{H}_k \right)^T,
	\end{split}
\end{align}
where $ \mathbf{F}^{\dagger} = \left( \mathbf{F}^H \mathbf{F} \right)^{-1} \mathbf{F}^H $. Note that the Wirtinger derivative of $ J^{F}_k  $ with respect to $ \mathbf{F}^* $ is zero, i.e., $ \nabla_{\mathbf{F}^*} J^{F}_k = \nabla_{\mathbf{F}^H} J^{F}_k = \mathbf{0} $.

\setcounter{equation}{0}
\renewcommand{\theequation}{B.\arabic{equation}}
\section{SDR-C Scheme} \label{xB}

\noindent{\textit{B.1. Optimization of} $\mathbf{F}$}

Assuming that $ \left\lbrace \mathbf{w}_k \right\rbrace^K_{k=1} $ and $ \mathbf{m} $ are known, notice that we can express $ \mathbf{F} \mathbf{m} = \mathbf{P} \mathbf{f} $, where $ \mathbf{P} = \mathbf{m}^T \otimes \mathbf{I} $ and $ \mathbf{f} = \mathrm{vec} \left( \mathbf{F} \right) $. Furthermore, if we assign $ t = \min_{k \in \mathcal{K}} \frac{\left| \mathbf{w}^H_k \mathbf{H}_k \mathbf{F} \mathbf{m} \right|^2} {{\sigma}^2 P^\mathrm{max}_\mathrm{rx}} $, then $ \mathcal{P}^{\mathrm{hyb}}_1 $ in (\ref{e4}) can be equivalently expressed as,
% Equation 14
\begin{subequations} \label{e14}
	\begin{align}
	% Objective 14a
	\mathcal{P}^{\mathrm{hyb}}_1: & \max_{\substack{ t, \mathbf{F} }} & & t \label{e14a}
	\\
	\vspace{-0.2cm}
	% Constraint 14b
	& ~~ \mathrm{s.t.} & & \left| \mathbf{w}^H_k \mathbf{H}_k \mathbf{P} \mathbf{f} \right|^2 \geq t, \label{e14b}
	\\
	% Constraint 14c
	& & & \left\| \mathbf{P} \mathbf{f} \right\|^2_2 = P^\mathrm{max}_\mathrm{tx}, \label{e14c}
	\\
	% Constraint 14d
	& & & \left[ \mathbf{f} \right]_n \in \mathcal{F}, n \in \mathcal{N}, \label{e14d}
	\end{align}
\end{subequations}
where $ \mathcal{N} = \left\lbrace 1, 2, \dots, N_\mathrm{tx} N^\mathrm{RF}_\mathrm{tx} \right\rbrace $. In (\ref{e14}), realize that $ \left\| \mathbf{P} \mathbf{f} \right\|^2_2 = \mathrm{Tr} \left( \mathbf{X} \mathbf{D} \right) $, with $ \mathbf{X} = \mathbf{P}^H \mathbf{P} $ and $\mathbf{D} = \mathbf{f}\mathbf{f}^H $. Also, $ \left[ \mathbf{D} \right]_{n,n} = \delta_\mathrm{tx} $ since $ \left[ \mathbf{f} \right]_n \in \mathcal{F} $. By noticing that $ \left| \mathbf{w}^H_k \mathbf{H}_k \mathbf{P} \mathbf{f} \right|^2 = \mathrm{Tr} \left( \mathbf{R}_k \mathbf{D} \right) $, with $ \mathbf{R}_k = \mathbf{P}^H \mathbf{H}^H_k \mathbf{w}_k \mathbf{w}^H_k \mathbf{H}_k \mathbf{P} $, (\ref{e14}) can be recast in its SDR form as shown in (\ref{e15})
% Equation 15
\begin{subequations} \label{e15}
	\begin{align}
	% Objective 15a
	\mathcal{P}^{\mathrm{hyb}}_{\mathrm{SDR},1}: & \max_{\substack{ t, \mathbf{D}}} & & t \label{e15a}
	\\
	\vspace{-0.2cm}
	% Constraint 15b
	& ~~ \mathrm{s.t.} & & \mathrm{Tr} \left\lbrace \mathbf{R}_k \mathbf{D} \right\rbrace \geq t, \label{e15b}
	\\
	% Constraint 15c
	& & & \left[ \mathbf{D} \right]_{n,n} = \delta_\mathrm{tx},  n \in \mathcal{N}, \label{e15c}
	\\
	% Constraint 15d
	& & & \mathbf{D} \succcurlyeq \mathbf{0}, \label{e15d}
	\end{align}
\end{subequations}
where the constraint $ \mathrm{rank} \left( \mathbf{D} \right) = 1 $ has been dropped. Also, (\ref{e15d}) enforces $ \mathbf{D} $ to be Hermitian positive semidefinite (PSD). Note that (\ref{e15d}) is linear in the PSD domain, and thus can be effectively approached by optimization solvers such as \texttt{SDPT3}. Upon obtaining $ \mathbf{D} $, $ \mathbf{f} $ is recovered in three stages.

\underline{\textit{Stage 1}:}
Observe that any element $\left( n_1, n_2 \right) $ of matrix $ \mathbf{D} $ can be represented as $ \left[ \mathbf{D} \right]_{n_1,n_2}  = \left[ \mathbf{f} \right]_{n_1} \left[ \mathbf{f} \right]^{*}_{n_2} $. Now, let us define a vector $ \mathbf{u} \in \mathbb{C}^{ N^\mathrm{RF}_\mathrm{tx} \times 1} $ such that $ \left\| \mathbf{u}\right\|^2_2 = \mathbf{u}^H \mathbf{u} = 1 $. Thus, we can express $ \left[ \mathbf{D} \right]_{n_1,n_2} $ in terms of $ \mathbf{u} $, i.e., $ \left[ \mathbf{D} \right]_{n_1,n_2}  = \left( \left[ \mathbf{f} \right]_{n_1} \mathbf{u}^T \right) \left( \left[ \mathbf{f} \right]^{*}_{n_2} \mathbf{u}^{*} \right) $. Assuming that $ \mathbf{q}_{n} = \left[ \mathbf{f} \right]_{n} \mathbf{u} $, $ \mathbf{D} $ can be recast as $ \mathbf{D} = \mathbf{Q}^T \mathbf{Q}^{*} $ with $ \mathbf{Q} = \left[ \mathbf{q}_1, \mathbf{q}_2, \dots, \mathbf{q}_{N_\mathrm{tx} N^\mathrm{RF}_\mathrm{tx}} \right] $. 

\underline{\textit{Stage 2}:}
If the solution returned by $ \mathcal{P}^{\mathrm{hyb}}_{\mathrm{SDR},1} $ is denoted by $ \widehat{\mathbf{D}} $. Then, via Cholesky decomposition we can obtain $ \widehat{\mathbf{D}} = \widehat{\mathbf{Q}}^T \widehat{\mathbf{Q}}^{*} $, where $ \widehat{\mathbf{Q}} = \left[ \widehat{\mathbf{q}}_1, \widehat{\mathbf{q}}_2, \dots, \widehat{\mathbf{q}}_{N_\mathrm{tx} N^\mathrm{RF}_\mathrm{tx}} \right]  $. In the previous stage, the premise was that each $ \mathbf{q}_n $ could be obtained from the same $ \mathbf{u} $, since $ \mathbf{q}_n = \left[ \mathbf{f} \right]_n \mathbf{u} $. However, we cannot guarantee that every $ \widehat{\mathbf{q}}_n $ in $ \widehat{\mathbf{D}} $ has the same stem $ \widehat{\mathbf{u}} $. Although we have found $ \widehat{\mathbf{D}} $, $ \mathbf{f} $ and $ \widehat{\mathbf{u}} $ remain unknown. 

\underline{\textit{Stage 3}:} 
The objective is to find some $ \widehat{\mathbf{u}} $ such that it originates the least error in the 2-norm sense, i.e., 
% Equation 16
\begin{subequations} \label{e16}
	\begin{align}
	% Objective 16a
	\mathcal{P}^{\mathrm{hyb}}_{\mathrm{LS},1}: & \min_{
													\substack{ 
																\widehat{\mathbf{u}}, 
																\left[ \mathbf{f} \right]_{n},
																\forall n \in \mathcal{N}
											 				 }
								   				   } & &
	{ 
		\sum^{N_\mathrm{tx} N^\mathrm{RF}_\mathrm{tx}}_{n=1} \left\| \widehat{\mathbf{q}}_n - \left[ \mathbf{f} \right]_n  \widehat{\mathbf{u}} \right\|^2_2
	} \label{e16a}
	\\
	\vspace{-0.2cm}
	% Constraint 16b
	& ~~~~~ \mathrm{s.t.} & & \left\| \widehat{\mathbf{u}} \right\|^2_2 = 1, \label{e16b}
	\\
	% Constraint 16c
	& & & \left[ \mathbf{f} \right]_n \in \mathcal{F}, n \in \mathcal{N}. \label{e16c}
	\end{align}
\end{subequations}

Minimizing simultaneously over both $ \widehat{\mathbf{q}}_n $ and $ \widehat{\mathbf{u}} $ is challenging. If we assume that $ \widehat{\mathbf{u}} $ is known such that (\ref{e8b}) is satisfied, then we are required to solve 
% Equation 17
\begin{subequations} \label{e17}
	\begin{align}
	% Objective 17a
	\widetilde{\mathcal{P}}^{\mathrm{hyb}}_{\mathrm{LS},1}: & \min_{
															\substack{ 
															\left[ \mathbf{f} \right]_{n},
															\forall n \in \mathcal{N}
														 		     }
											   			     } & &
	{ 
		\sum^{N_\mathrm{tx} N^\mathrm{RF}_\mathrm{tx}}_{n=1} \left\| \widehat{\mathbf{q}}_n - \left[ \mathbf{f} \right]_n  \widehat{\mathbf{u}} \right\|^2_2
	} \label{e17a}
	\\
	\vspace{-0.2cm}
	% Constraint 17b
	& ~~~~ \mathrm{s.t.} & & \left[ \mathbf{f} \right]_n \in \mathcal{F}, n \in \mathcal{N}. \label{e17b}
	\end{align}
\end{subequations}

By expanding (\ref{e17a}), we realize that $ \left\| \widehat{\mathbf{q}}_n - \left[ \mathbf{f} \right]_n  \widehat{\mathbf{u}} \right\|^2_2 = \widehat{\mathbf{q}}^H_n \widehat{\mathbf{q}}_n - 2 \mathfrak{Re} \left( \left[ \mathbf{f} \right]_n \widehat{\mathbf{q}}^H_n \widehat{\mathbf{u}} \right) + \left| \left[ \mathbf{f} \right]_n \right|^2 \widehat{\mathbf{u}}^H \widehat{\mathbf{u}} $. Thus, (\ref{e17}) is
\begin{subequations} \label{e18}
	\begin{align}
	% Objective 18a
	\widetilde{\mathcal{P}}^{\mathrm{hyb}}_{\mathrm{LS},1}: & \max_{
												\substack{ 
															\left[ \mathbf{f} \right]_{n},
															\forall n \in \mathcal{N}
														 }
											   } & & 
	{ 
		\sum^{N_\mathrm{tx} N^\mathrm{RF}_\mathrm{tx}}_{n=1} \mathfrak{Re} \left( \left[ \mathbf{f} \right]_n \widehat{\mathbf{q}}^H_n \widehat{\mathbf{u}} \right)
	} \label{e18a}
	\\
	\vspace{-0.2cm}
	% Constraint 18b
	& ~~~~ \mathrm{s.t.} & & \left[ \mathbf{f} \right]_n \in \mathcal{F}, n \in \mathcal{N}. \label{e18b}
	\end{align}
\end{subequations}

Note that (\ref{e18}) can be decomposed into $ N_\mathrm{tx} N^\mathrm{RF}_\mathrm{tx} $ independent sub-problems. Thus, since $ z_n = \widehat{\mathbf{q}}^H_n \widehat{\mathbf{u}} $ is known, we need to select $ \left[ \mathbf{f} \right]_n $ such that the real part of (\ref{e18a}) is maximized. This is equivalent to choosing $ \left[ \mathbf{f} \right]_n $ with the closest phase to $ z^{*}_n $. After finding $ \mathbf{f} $, it can be reshaped to obtain $ \mathbf{F} $. \\

\noindent{\textit{B.2. Optimization of} $\mathbf{m}$}

We assume herein that $ \mathbf{F} $ and $ \left\lbrace \mathbf{w}_k \right\rbrace^K_{k=1} $ are known. Thus, the SDR form of $ \mathcal{P}^{\mathrm{hyb}}_2 $ is given by
% Equation 20
\begin{subequations} \label{e20}
	\begin{align}
	% Objective 20a
	\mathcal{P}^{\mathrm{hyb}}_{\mathrm{SDR},2}: & \max_{\substack{ t, \mathbf{M}}} & & t \label{e20a}
	\\
	\vspace{-0.2cm}
	% Constraint 20b
	& ~~ \mathrm{s.t.} & & \mathrm{Tr} \left( \mathbf{Z}_k \mathbf{M} \right) \geq t, \label{e20b}
	\\
	% Constraint 20c
	& & & \mathrm{Tr} \left( \mathbf{Y} \mathbf{M} \right) = P^\mathrm{max}_\mathrm{tx}, \label{e20c}
	\\
	% Constraint 20d
	& & & \mathbf{M} \succcurlyeq \mathbf{0}, \label{e20d}
	\end{align}
\end{subequations}
where $ \mathbf{Y} = \mathbf{F}^H \mathbf{F} $,  $ \mathbf{Z}_k = \mathbf{F}^H \mathbf{H}^H_k \mathbf{w}_k \mathbf{w}^H_k \mathbf{H}_k \mathbf{F} $ and $ \mathbf{M} = \mathbf{m} \mathbf{m}^H $.\\

\noindent{\textit{B.3. Optimization of} $\mathbf{w}_k$}

Now, we assume that $ \mathbf{F} $ and $ \mathbf{m} $ are given. Therefore, SDR form of $ \mathcal{P}^{\mathrm{hyb}}_3 $ is
% Equation 22
\begin{subequations} \label{e22}
	\begin{align}
	% Objective 22a
	\mathcal{P}^{\mathrm{hyb}}_{\mathrm{SDR},3}: & \max_{\substack{ t,\left\lbrace \mathbf{W}_k \right\rbrace^K_{k = 1}}} & & t \label{e22a}
	\\
	\vspace{-0.2cm}
	% Constraint 22b
	& ~~ \mathrm{s.t.} & & \mathrm{Tr} \left( \mathbf{C}_k \mathbf{W}_k \right) \geq t, \label{e22b}
	\\
	% Constraint 22c
	& & & \mathrm{Tr} \left( \mathbf{W}_k \right) = P^\mathrm{max}_\mathrm{rx}, \label{e22c}
	\\
	% Constraint 22d
	& & & \mathbf{W}_k \succcurlyeq \mathbf{0}, k \in \mathcal{K}, \label{e22d}
	\end{align}
\end{subequations}
where $ \mathbf{W}_k = \mathbf{w}_k \mathbf{w}^H_k $ and $ \mathbf{C}_k = \mathbf{H}_k \mathbf{F} \mathbf{m} \mathbf{m}^H \mathbf{F}^H \mathbf{H}^H_k $. The problems $ \mathcal{P}^{\mathrm{hyb}}_{\mathrm{SDR},1} $, $ \mathcal{P}^{\mathrm{hyb}}_{\mathrm{SDR},2} $ and $ \mathcal{P}^{\mathrm{hyb}}_{\mathrm{SDR},3} $ can be straightforwardly recast as linear programs and can therefore be efficiently solved by numerical solvers. In our case, we employed \texttt{CVX} and \texttt{SDPT3}.

\end{appendices}

\bibliographystyle{IEEEtran}
\bibliography{ref}

\end{document}